\documentclass[10pt,letterpaper]{article}
\usepackage[margin=1in]{geometry}
\usepackage{amsmath,amssymb,amsthm}
\usepackage{booktabs}
\usepackage{microtype}
\usepackage{graphicx}

\newtheorem{proposition}{Proposition}
\newtheorem{theorem}{Theorem}
\newtheorem{corollary}[proposition]{Corollary}
\newtheorem{lemma}[proposition]{Lemma}
\theoremstyle{definition}
\newtheorem{assumption}{Assumption}
\theoremstyle{remark}

\usepackage[colorinlistoftodos]{todonotes}
\usepackage[colorlinks=true, allcolors=blue]{hyperref}
\usepackage[normalem]{ulem}

\usepackage{titlesec}

\titlespacing{\paragraph}{0pt}{2.5ex plus 1ex minus .2ex}{1em}
\newcommand{\scomment}[1]{}
\newcommand{\srcomment}[1]{}
\newcommand{\jcomment}[1]{}

\DeclareMathOperator{\tr}{tr}

\newcommand{\E}{\mathbb{E}}

\title{Bias in Method-of-Moments Heritability Estimation\\
Under Model Misspecification}

\author{%
  Jonathan Gal\textsuperscript{1}\qquad
  Sriram Sankararaman\textsuperscript{2}\qquad
  Saharon Rosset\textsuperscript{1}\\[1em]
  \normalsize
  \textsuperscript{1}Department of Statistics and Operations Research,
  School of Mathematical Sciences,\\
  \normalsize Tel Aviv University, Tel Aviv, Israel\\[0.5em]
  \normalsize
  \textsuperscript{2}Departments of Computer Science,
  Human Genetics,\\
  \normalsize and Computational Medicine,\\
  \normalsize University of California, Los Angeles,
  Los Angeles, CA, USA
}
\date{}

\DeclareMathOperator{\offop}{off}
\DeclareMathOperator{\diagop}{diag}

\newenvironment{restated}[2]
  {\begin{trivlist}
   \item[\hskip\labelsep{\bfseries #1~#2.}]
   \itshape}
  {\end{trivlist}}

\begin{document}

\maketitle
\thispagestyle{empty}

\begin{abstract}
\normalsize
Method-of-Moments (MoM) based approaches for estimation of variance components --- heritability and beyond --- are central in computational genetics due to their flexibility in addressing diverse and complex models while minimizing assumptions, and ease of computation and approximation in biobank-scale data. MoM are closely related to other important methods in this area like LD-score regression. 
A fundamental, highly debated question relates to the bias inherent in MoM estimates of variance components when the fitted model is misspecified: since the true underlying model is unknown, the fitted model may be missing some of the components from the true model or contain superfluous ones. 
We offer a general mathematical formulation of this bias and investigate its properties. This leads to novel insights about the sign and the magnitude of the bias in settings of interest and allows extensions and re-interpretation of previous MoM results in the literature. 
One case study we pursue is in estimating gene-environment interaction variance components (GxE). We prove that when omitting environment-specific noise components (NxE), the resulting GxE estimates are strictly upward biased, even when no true interaction exists.
In another case study, we consider heritability estimation, when the genetic variance components are incorrectly partitioned (such as using a single genome-wide component instead of chromosome-specific components). We derive explicit expressions for the bias in the estimate of genome-wide heritability and its dependence on LD scores within and between components. For example, if all components have the same average LD score, the misspecified heritability estimate is still unbiased.
\end{abstract}

\clearpage
\setcounter{page}{1}

\section{Introduction}
A central goal of modern human genetics is to understand the genetic architecture of complex traits: how much of trait variation is attributable to genetic variation ({\em heritability}) and how this heritability varies across genomic and environmental context. 
Given that the genetic basis of most complex traits is attributed to a large number of variants of small effect~\cite{visscher2017}, answering questions about genetic architecture requires modeling the joint contribution of all variants rather than individual loci. 
Linear mixed models (LMMs), and the variance components that parameterize them, are a key tool in this effort~\cite{yang2010,gcta2011}. In these models, the phenotypic covariance among individuals is written as a sum of variance components, each scaling a known kernel (a matrix of similarities between individuals).
The additive genetic component, with a kernel given by the genetic relatedness matrix (GRM), yields
SNP heritability \cite{yang2010}; partitioning SNPs into multiple kernels by chromosome, minor allele
frequency (MAF), levels of linkage disequilibrium (LD) or functional annotation localizes heritability across the genome
\cite{yang2015,finucane2015,rhemc}; and kernels that interact genotypes with environmental variables
quantify gene--environment (G$\times$E) interactions \cite{robinson2017,gxemm,genie}. 
The variance components are thus directly interpretable as quantities of biological interest.

The classical approach to estimating variance components is restricted maximum likelihood (REML)
\cite{gcta2011,loh2015}. While REML estimates are statistically efficient (assuming the model is correctly specified), the cost of computing the REML estimates makes them impractical at biobank scale even with a number of computational advances~\cite{fastlmm,gemma,boltlmm,loh2015}. Method-of-moments (MoM)
estimators, of which Haseman--Elston (HE) regression \cite{haseman1972} is the best-known example, instead
match the empirical moments of the phenotype to those implied by the model. MoM estimators are more robust to model misspecification (typically requiring only the first two moments of the phenotype). Combined with recent randomized approximations, MoM estimators can scale to
millions of individuals and many kernels \cite{rhe,rhemc,genie}. Importantly, MoM estimators can be adapted to work with GWAS summary
statistics~\cite{zhou2017,ldsc,finucane2015,speed2019}. These properties have made moment-based estimation the dominant approach for heritability analyses.

The validity of MoM estimates, however, rests on the correctness of the fitted model. Because the true model is unknown, any fitted model may omit components of the true
covariance or include superfluous ones leading to potential biases in estimates of the remaining components. It is now understood that estimates of SNP heritability depend strongly on the assumed
relationship between per-SNP effects, MAF and LD \cite{speed2012,speed2017,evans2018}: the GCTA
model~\cite{gcta2011,yang2010} (which also underlies standard LDSC~\cite{ldsc}) makes different assumptions about this relationship relative to the LDAK model~\cite{speed2012,speed2017}. The two models give substantially different estimates of SNP heritability
and of the enrichment of functional categories, and the question of which model better describes real traits has been the subject of an extended debate
\cite{speed2017,gazal2017,speed2019,speed2020}. 
Thus, it is important to understand how heritability estimates are affected when the assumptions of the true generative model differ from those of the fitted one. While approaches that stratify SNPs by MAF and LD have been shown empirically to reduce this sensitivity~\cite{yang2015,evans2018,hou2019}, a general characterization of the bias of MoM under arbitrary misspecified partitions has been lacking.

Another setting where these concerns arise is in the estimation of G$\times$E heritability: a topic of intense recent interest motivated by the availability of diverse environmental exposures, in addition to genotype-phenotype data, in biobanks~\cite{robinson2017,gxemm,gplemma,genie}. Environmental variables, however, can
modulate not only genetic effects but also the residual variance, so that the noise variance differs
across environments (termed noise-by-environment, or N$\times$E, heterogeneity). 
Models that fit G$\times$E kernels without accounting for N$\times$E heterogeneity can attribute environment-dependent noise to
genetic interactions. This produces spurious G$\times$E heritability even when no interaction exists~\cite{gxemm,genie}. 
Conversely, the common practice of fitting
only an additive genetic component when G$\times$E is present can bias estimates of additive heritability
in either direction, depending on how the environment aligns with genetic relatedness. In each of these
settings, the bias has so far been characterized empirically or under restrictive assumptions~\cite{gxemm}.

\noindent
\textbf{Our contributions}: 
Here we present a general, exact formulation of the conditional bias of MoM variance-component
estimates when the fitted model omits components of the true covariance. We show that the bias is
additive across omitted components and is determined by the Frobenius alignment of the omitted kernels
with the fitted ones. We further show that, when the fitted genetic kernels have diagonal ``twins'' in the
fit, the genetic estimates are determined entirely by the off-diagonal entries of the phenotypic
covariance, so that omitted diagonal (noise) components cannot bias them. Applying this framework, we
prove that omitting N$\times$E strictly inflates G$\times$E variance estimates, even when no true
interaction exists. We characterize the sign and magnitude of additive and interaction bias with
multiple, correlated environments. Finally, we derive an explicit LD-score expression for the bias of
genome-wide heritability when genetic components are incorrectly pooled, which clarifies when the GCTA-
and LDAK-type disagreements can and cannot arise. Simulations confirm that empirical estimates agree
with the exact predictions.

\section{Methods}
Let $y\in\mathbb R^N$ be a centered phenotype vector, $X\in\mathbb R^{N\times M}$ a column-standardized genotype matrix of $M$ SNPs, and $E\in\mathbb R^{N\times L}$ a matrix of $L$ column-standardized environmental variables. Throughout, we condition on the
genotypes $X$ and, if present in the data, environmental variables $E$, treating the associated
kernels as fixed, and do not include fixed-effect covariates. We consider the general covariance model
\[
\Sigma=\operatorname{Cov}(y\mid X,E)
=\sum_{j=1}^p\theta_jK_j,
\]
where $K_1,\ldots,K_p$ are known symmetric matrices.
For variance components, $\theta_j=\sigma_j^2$. For example, the GRM $K_g=XX^\top/M$ corresponds to an additive genetic variance component $\sigma^2_g$,
while $I_N$ corresponds to an independent residual component $\sigma^2_e$.

The standard method-of-moments (MoM) \cite{zhou2017} estimator matches the observed matrix $yy^\top$, including its
diagonal, to the modeled covariance in Frobenius norm:
$
\widehat\theta
=\arg\min_{a\in\mathbb R^p}
\left\|yy^\top-\sum_{j=1}^p a_jK_j\right\|_F^2.
$
Writing $\langle A,B\rangle_F=\operatorname{tr}(A^\top B)$,
the normal equations yield:
\[
\widehat\theta=T^{-1}q,
\qquad
T_{rs}=\langle K_r,K_s\rangle_F
=\operatorname{tr}(K_rK_s),
\qquad
q_r=\langle K_r,yy^\top\rangle_F=y^\top K_r y,
\]
provided the fitted kernels are linearly independent.
Under correct specification,
$\mathbb E[q_r\mid X,E]=\operatorname{tr}(K_r\Sigma)$,
so the estimator is conditionally unbiased:
$\mathbb E[\widehat\theta\mid X,E]
=T^{-1}\mathbb E[q\mid X,E]
=T^{-1}T\theta=\theta$.

Classical Haseman--Elston (HE) regression~\cite{haseman1972} instead regresses the phenotypic products $y_iy_j$ of
distinct pairs on their relatedness, using only the off-diagonal entries of $yy^\top$.
For $\{K_g,I_N\}$ the two estimators coincide when $\operatorname{diag}(K_g)=\mathbf 1$, which holds approximately for
standardized genotypes, and MoM is therefore commonly called HE regression. Section~\ref{sec:separation} makes this
exact for general kernel sets: the MoM genetic estimates equal the off-diagonal HE estimates
whenever the fitted diagonal kernels satisfy the twin condition. Our analysis concerns the
unconstrained, full-matrix MoM estimator with exact kernel inner products.

\paragraph{Heritability.}
For each environmental variable $\ell=1,\ldots,L$, let $D_\ell=\operatorname{diag}(E_{\cdot\ell})$.
Per-SNP interaction effects $\alpha_\ell\sim(0,\sigma^2_{gxe,\ell}M^{-1}I_M)$ in
$y=X\beta+\sum_{\ell=1}^LD_\ell X\alpha_\ell+\cdots$ contribute the G$\times$E kernels
$K_{gxe,\ell}=D_\ell K_gD_\ell$. For a genetic component $j$ ($g$, or $gxe,\ell$) we define its
heritability as its share of the total phenotypic variance,
\[
h^2_j=\frac{\sigma^2_j\operatorname{tr}(K_j)}{\operatorname{tr}(\Sigma)},
\qquad\text{so}\qquad
h^2_g=\frac{\sigma^2_g\operatorname{tr}(K_g)}{\operatorname{tr}(\Sigma)},
\qquad
h^2_{gxe,\ell}=\frac{\sigma^2_{gxe,\ell}\operatorname{tr}(K_{gxe,\ell})}{\operatorname{tr}(\Sigma)},
\qquad
h^2_{gxe}=\sum_{\ell=1}^Lh^2_{gxe,\ell},
\]
where $\operatorname{tr}(K_j)/N$ is the average variance contributed per unit of $\sigma^2_j$
and $\operatorname{tr}(\Sigma)/N$ is the average phenotypic variance. With standardized genotypes
$\operatorname{tr}(K_g)=N$, and with standardized environmental variables and
$\operatorname{diag}(K_g)\approx\mathbf 1$,
$\operatorname{tr}(K_{gxe,\ell})=\sum_iK_{g,ii}E_{i\ell}^2\approx N$. Heritabilities are estimated by plugging in the fitted components,
$
\widehat h^2_j
=\frac{\widehat\sigma^2_j\operatorname{tr}(K_j)}{\sum_{r\in\mathcal F}\widehat\theta_r\operatorname{tr}(K_r)}$,
where $\mathcal F$ is the index set of fitted kernels.
\subsection{Exact additive bias under covariance misspecification}

Let $J=\{1,\ldots,p\}$ index the kernels in the true covariance
model $\Sigma=\sum_{j\in J}\theta_jK_j$.
Suppose we fit only the kernels indexed by a nonempty subset
$\mathcal F\subseteq J$, and let $\mathcal O=J\setminus\mathcal F$
denote the omitted indices. Define
\[
\theta_{\mathcal F}=(\theta_r)_{r\in\mathcal F}
\in\mathbb R^{|\mathcal F|},
\qquad
q_{\mathcal F}=(y^\top K_r y)_{r\in\mathcal F}
\in\mathbb R^{|\mathcal F|},
\quad
T_{\mathcal F}
=\bigl(\langle K_r,K_s\rangle_F\bigr)_{r,s\in\mathcal F}
\in\mathbb R^{|\mathcal F|\times|\mathcal F|}.
\]
Here $\theta_{\mathcal F}$ contains the true coefficients of just
the fitted kernels. The fitted MoM estimator is
$\widehat\theta_{\mathcal F}=T_{\mathcal F}^{-1}q_{\mathcal F}$.

\begin{theorem}[Additivity of underspecification bias]
\label{thm:additive-bias}
Suppose $\E[y\mid X,E]=0$ and the fitted kernels are linearly
independent. For each omitted index $j\in\mathcal O$, define
$
v_j
=\bigl(\langle K_r,K_j\rangle_F\bigr)_{r\in\mathcal F}
\in\mathbb R^{|\mathcal F|},
$
whose entries are the Frobenius inner products of $K_j$
with the fitted kernels. Then
\[
\boxed{
\operatorname{Bias}(\widehat\theta_{\mathcal F}\mid X,E)
:=\E[\widehat\theta_{\mathcal F}\mid X,E]-\theta_{\mathcal F}
=\sum_{j\in\mathcal O}\theta_jT_{\mathcal F}^{-1}v_j.
}
\]
\end{theorem}

\begin{proof}
For every $r\in\mathcal F$, the conditional mean-zero
assumption gives
\[
\E[y^\top K_r y\mid X,E]
=\operatorname{tr}(K_r\Sigma)
=\sum_{s\in\mathcal F}\theta_s\langle K_r,K_s\rangle_F
+\sum_{j\in\mathcal O}\theta_j\langle K_r,K_j\rangle_F.
\]
Collecting these identities yields
\[
\E[q_{\mathcal F}\mid X,E]
=T_{\mathcal F}\theta_{\mathcal F}
+\sum_{j\in\mathcal O}\theta_jv_j.
\]
Since $T_{\mathcal F}$ is fixed conditional on $(X,E)$
and invertible,
\[
\begin{aligned}
\E[\widehat\theta_{\mathcal F}\mid X,E]-\theta_{\mathcal F}
=T_{\mathcal F}^{-1}\E[q_{\mathcal F}\mid X,E]
  -\theta_{\mathcal F}
=T_{\mathcal F}^{-1}
  \left(T_{\mathcal F}\theta_{\mathcal F}
  +\sum_{j\in\mathcal O}\theta_jv_j\right)
  -\theta_{\mathcal F}
=\sum_{j\in\mathcal O}\theta_jT_{\mathcal F}^{-1}v_j,
\end{aligned}
\]
which proves the claim.
\end{proof}

Thus, each omitted component contributes additively to the bias linearly through its coefficient $\theta_j$ (typically a variance component $\sigma_j^2$), the Frobenius alignment of its kernel with the fitted kernels, and the structure of $T_{\mathcal F}^{-1}$.
\begin{corollary}[Unbiasedness under overspecification]
\label{cor:overspecification}
If $\mathcal F\supseteq J$ and the fitted kernels are linearly independent, then MoM remains unbiased: $\E[\widehat\theta_r\mid X,E]=\theta_r$ for $r\in J$ and $0$ for $r\in\mathcal F\setminus J$.
\end{corollary}
A proof is given in Supplementary Section~\ref{sec:overspecification}.

\subsection{Diagonal-off-diagonal separation and connection to HE regression}

\label{sec:separation}
We show that when the fitted kernel set is ``complete'' in a sense made precise below, MoM
decomposes into two successive steps, and the components that matter for heritability are
estimated entirely in the first.

Split the fitted kernels as $\mathcal F=\mathcal G\cup\mathcal H$, where every kernel in $\mathcal H$
is diagonal (e.g.\ $I_N$, or $K_{nxe,\ell}=D_\ell^2$) and $\mathcal G$ contains the remaining kernels
(e.g.\ $K_g$, $K_{gxe,\ell}$), which we call genetic. Write $\operatorname{off}(K)=K-\operatorname{diag}(K)$
for the off-diagonal part of a matrix, $y\circ y=(y_i^2)_{i=1}^N$, and
$\Phi=[\operatorname{diag}(H)]_{H\in\mathcal H}\in\mathbb R^{N\times|\mathcal H|}$ for the matrix whose
columns are the diagonals of the fitted diagonal kernels.

\begin{assumption}[Twin condition]
\label{ass:twin}
There exists a matrix
$\Lambda\in\mathbb R^{|\mathcal H|\times|\mathcal G|}$
such that, for every $G_a\in\mathcal G$,
\[
\operatorname{diag}(G_a)=\Phi\Lambda_{\cdot a}.
\]
That is, the diagonal of every fitted genetic kernel lies
in the span of the fitted diagonal kernels, in particular when each genetic kernel has a diagonal ``twin'' in the fit.
\end{assumption}
For example, for $\mathcal F=\{K_g,I_N\}$ the condition
holds exactly when $\operatorname{diag}(K_g)=\mathbf 1$, and approximately for standardized genotypes,
where $\operatorname{diag}(K_g)\approx\mathbf 1$. For a G$\times$E kernel,
$\operatorname{diag}(K_{gxe,\ell})=D_\ell^2\operatorname{diag}(K_g)$, so with
$\operatorname{diag}(K_g)=\mathbf 1$ its twin is the N$\times$E kernel $K_{nxe,\ell}=D_\ell^2$.

\begin{theorem}[Diagonal--off-diagonal separation]
\label{thm:separation}
Suppose the fitted kernels are linearly independent and satisfy the twin condition. Then the MoM
estimates of the genetic components are obtained from the off-diagonal entries of $yy^\top$ alone,
\[
\boxed{
\widehat\theta_{\mathcal G}=(T^{\mathrm{off}})^{-1}q^{\mathrm{off}},
}
\qquad
T^{\mathrm{off}}_{ab}=\langle\operatorname{off}(G_a),\operatorname{off}(G_b)\rangle_F,
\qquad
q^{\mathrm{off}}_a= y^\top \operatorname{off}(G_a)y.
\]
and the diagonal components are then estimated from the diagonal residuals,
\[
\widehat\theta_{\mathcal H}=(\Phi^\top\Phi)^{-1}\Phi^\top\bigl(y\circ y-\Phi\Lambda\widehat\theta_{\mathcal G}\bigr)
=(\Phi^\top\Phi)^{-1}\Phi^\top(y\circ y)-\Lambda\widehat\theta_{\mathcal G}.
\]
\end{theorem}

The proof (Supplementary Section~\ref{sec:separation-proof}) uses that diagonal and off-diagonal matrices are Frobenius-orthogonal,
so the MoM objective splits into an off-diagonal term that involves only $\theta_{\mathcal G}$ and a diagonal
term; Under the twin condition, minimizing the diagonal term over
$\theta_{\mathcal H}$ gives the same minimum for every value of
$\theta_{\mathcal G}$, so the genetic estimates are determined
entirely by the off-diagonal term. Thus,
$\widehat\theta_{\mathcal G}$ coincides with the corresponding
multicomponent off-diagonal HE estimator.

\begin{corollary}[Omitted diagonal kernels]
\label{cor:omitted-diag}
Suppose the fitted kernel set $\mathcal G\cup\mathcal H$ satisfies the twin condition, and denote by $\widehat\theta_{\mathcal G}\bigl(\mathcal G\cup\mathcal H\bigr)$ the MoM genetic estimator based on this fitted set. Then adding any diagonal kernels to the fit
leaves the genetic estimates unchanged: for every phenotype vector $y$, whatever the true model,
\[
\boxed{
\widehat\theta_{\mathcal G}\bigl(\mathcal G\cup\mathcal H\bigr)=\widehat\theta_{\mathcal G}\bigl(\mathcal G\cup\mathcal H'\bigr)
\quad\text{for all diagonal }\mathcal H'\supseteq\mathcal H.
}
\]
\end{corollary}
Thus, omitting a diagonal component leaves the genetic
estimator, and hence its distribution, unchanged. For example, when $\operatorname{diag}(K_g)=\mathbf 1$,
fitting $\{K_g,I_N\}$ or $\{K_g,K_{nxe},I_N\}$ gives
exactly the same $\widehat\sigma_g^2$, whether or not the
true model contains N$\times$E.
Together with Theorem~\ref{thm:additive-bias}, this gives 
$\operatorname{Bias}(\widehat\theta_{\mathcal G}\mid X,E)=0$
whenever all omitted kernels are diagonal.

\subsection{Direct implications for heritability estimates}
\label{sec:h2}
Heritability is a ratio, so a biased component could in principle distort every heritability estimate
through the denominator $\sum_r\widehat\theta_r\operatorname{tr}(K_r)$. The following identity shows that it in fact
does not.

\begin{lemma}[Trace conservation]
\label{lemma:trace}
    Suppose $K_r=I_N$ for some $r\in\mathcal F$ and write the true covariance as
$\Sigma=\sum_{j\in\mathcal F}\theta_jK_j+\Delta$, where $\Delta=\sum_{j\in\mathcal O}\theta_jK_j$ is the
omitted covariance. With $B_j=\E[\widehat\theta_j\mid X,E]-\theta_j$ the bias of fitted component $j$,
\[
\sum_{j\in\mathcal F}\operatorname{tr}(K_j)B_j=\operatorname{tr}(\Delta),
\qquad\text{equivalently}\qquad
\sum_{j\in\mathcal F}\operatorname{tr}(K_j)\,\E[\widehat\theta_j\mid X,E]=\operatorname{tr}(\Sigma).
\]
\end{lemma}

The proof is given in Supplementary Section \ref{sec:trace-proof}.
The same row of the normal equations $T_{\mathcal F}\widehat\theta_{\mathcal F}=q_{\mathcal F}$ gives the identity $\sum_{j\in\mathcal F}\operatorname{tr}(K_j)\widehat\theta_j=y^\top y$.\\ MoM therefore
preserves total phenotypic variance on the trace scale: misspecification only redistributes that variance
among the fitted components, and the omitted variance $\operatorname{tr}(\Delta)$ is absorbed by them in
proportion to $\operatorname{tr}(K_j)$.

To compare heritabilities, we evaluate the plug-in heritability at the expected component estimates,
\[
h^{2,*}_j=\frac{(\theta_j+B_j)\operatorname{tr}(K_j)}{\sum_{r\in\mathcal F}(\theta_r+B_r)\operatorname{tr}(K_r)},
\]
and call $h^{2,*}_j-h^2_j$ the heritability \emph{pseudo-bias}: it is the error of the heritability implied by
the expected estimates rather than the expectation of the ratio $\widehat h^2_j$. By Lemma~\ref{lemma:trace} the denominator of $h^{2,*}_j$ equals $\operatorname{tr}(\Sigma)$, which gives:

\begin{corollary}[Heritability pseudo-bias]
\label{cor:h2-bias}
Suppose $K_r=I_N$ for some $r\in\mathcal F$. Then for every fitted genetic component $j$,
$\operatorname{sgn}(h^{2,*}_j-h^2_j)=\operatorname{sgn}(B_j)$ and
\[
\boxed{
h^{2,*}_j-h^2_j=\frac{\operatorname{tr}(K_j)}{\operatorname{tr}(\Sigma)}\,B_j
\;\overset{\text{(SA)}}{=}\;B_j,
}
\]
where the last equality holds under the standard assumptions (SA): $\operatorname{diag}(K_g)=\mathbf 1$,
$E_{\cdot\ell}^\top E_{\cdot\ell}=N$ and unit average phenotypic variance, which give
$\operatorname{tr}(K_g)=\operatorname{tr}(K_{gxe,\ell})=\operatorname{tr}(\Sigma)=N$. Thus, typically, the
heritability pseudo-bias of a genetic component equals its variance-component bias:
$h^{2,*}_g-h^2_g=B_g$ and $h^{2,*}_{gxe,\ell}-h^2_{gxe,\ell}=B_{gxe,\ell}$.
\end{corollary}

Thus, the residual-component bias does not enter any genetic heritability pseudo-bias, while its sign and
magnitude are exactly determined by the bias of the corresponding genetic variance component. Consequently, by Corollary~\ref{cor:omitted-diag}, as long as the twin condition holds, heritability estimates are
invariant to fitting any additional diagonal kernels. Indeed, by the replicate-wise identity above,
$\widehat h^2_j=\widehat\theta_j\operatorname{tr}(K_j)/y^\top y$, and neither $\widehat\theta_j$ nor $y^\top y$ changes when
diagonal kernels are added.

\section{Results}

\subsection{Bias in gene-environment interaction models}
\label{sec:single-environment}

Let $E\in\mathbb R^N$ be a standardized environmental variable,
so that $\mathbf1^\top E=0$ and $E^\top E=N$, and write
$D=\operatorname{diag}(E)$. Assume the true model
$y=X\beta+DX\alpha+D\delta+\varepsilon$,
where, conditional on $(X,E)$, the random effects are mutually
independent, have mean zero, and satisfy
\[
\begin{aligned}
\operatorname{Cov}(\beta\mid X,E)=\frac{\sigma_g^2}{M}I_M,
\quad
\operatorname{Cov}(\alpha\mid X,E)=\frac{\sigma_{gxe}^2}{M}I_M,
\quad
\operatorname{Cov}(\delta\mid X,E)=\sigma_{nxe}^2I_N,
\quad
\operatorname{Cov}(\varepsilon\mid X,E)=\sigma_e^2I_N.
\end{aligned}
\]
The conditional phenotypic covariance is
\[
\Sigma
=\sigma_g^2K_g+\sigma_{gxe}^2K_{gxe}
+\sigma_{nxe}^2K_{nxe}+\sigma_e^2I_N,
\; \mbox{where} \quad
K_g=\frac{XX^\top}{M},\;
K_{gxe}=DK_gD,\;
K_{nxe}=D^2.
\]
We refer to this as the $G+G\times E+N\times E$ model.
The $N\times E$ term represents residual variance that varies
with $E_i^2$.

For the explicit formulas below, assume
$\operatorname{diag}(K_g)=\mathbf1$ and that the fitted kernels
are linearly independent. Define the kernels after projecting
out $I_N$ by
$
\widetilde K_g=K_g-I_N,
\widetilde K_{gxe}=K_{gxe}-I_N$
with trace zero. In particular,
$\widetilde K_g=\operatorname{off}(K_g)$, whereas
$\widetilde K_{gxe}$ generally retains a nonzero diagonal.

\paragraph{Omitting environment-specific noise.}
Suppose MoM fits $\{K_g,K_{gxe},I_N\}$, omitting $K_{nxe}$  (for example, MEMMA or GPLEMMA~\cite{gplemma}).
Let $B_g$ and $B_{gxe}$ denote the conditional biases of
$\widehat\sigma_g^2$ and $\widehat\sigma_{gxe}^2$.

\begin{proposition}[Bias from omitted $N\times E$]
\label{prop:single-nxe-bias}
Under the assumptions above,
\[
\boxed{
B_{gxe}
=\sigma_{nxe}^2
\frac{
\|\widetilde K_g\|_F^2\,\|D^2-I_N\|_F^2
}{
\|\widetilde K_g\|_F^2\|\widetilde K_{gxe}\|_F^2
-\langle\widetilde K_g,\widetilde K_{gxe}\rangle_F^2
},
\quad
B_g
=-\frac{\langle\widetilde K_g,\widetilde K_{gxe}\rangle_F}
        {\|\widetilde K_g\|_F^2}\,B_{gxe}.
}
\]
If $\sigma_{nxe}^2>0$ and $D^2\neq I_N$, then
\[
\boxed{0<B_{gxe}\leq\sigma_{nxe}^2.
}
\]
\end{proposition}

The proof is given in Supplementary Section \ref{sec:omitted-nxe-proof}.
Thus, omitting environment-specific noise strictly inflates
the interaction variance estimate under these conditions,
even when the true interaction variance is zero. For fixed $(X,E)$, both biases are linear in $\sigma_{nxe}^2$
and independent of the true genetic variance components.
Unlike the interaction bias, the additive genetic bias can have
either sign. Specifically,
\[
\langle\widetilde K_g,\widetilde K_{gxe}\rangle_F
=\sum_{i\neq j}K_{g,ij}^2E_iE_j.
\]
The additive genetic bias is negative when pairs with
same-sign environmental values dominate this sum, weighted
by squared genetic relatedness, and positive when opposite-sign
pairs dominate it.

Under the additional normalization $\operatorname{tr}(\Sigma)=N$,
we have
$\operatorname{tr}(\Sigma)=\operatorname{tr}(K_g)
=\operatorname{tr}(K_{gxe})=N$.
Corollary~\ref{cor:h2-bias} therefore gives
\[
h_j^{2,*}-h_j^2=B_j,
\qquad j\in\{g,gxe\}.
\]
Hence the same signs and magnitudes hold for heritability pseudo-bias.
\paragraph{Connection to literature.}This result provides a theoretical explanation for the inflation
of $G\times E$ heritability observed in the GENIE simulations
when $N\times E$ is omitted~\cite{genie}.
Under the assumptions of Proposition~\ref{prop:single-nxe-bias},
we obtain an exact conditional bias formula and establish its
strict positivity for exact MoM; GENIE uses a randomized
approximation~\cite{rhe,rhemc}. Furthermore, GxEMM \cite[Appendix~A]{gxemm} derives an approximate MoM bias formula for discrete
environments under a pure-noise model, assuming approximately
equal group sizes and genotype--environment independence
.
Our general formulation gives exact conditional bias without
these restrictions, allowing nonzero genetic variance and
arbitrary fixed genotypes and environments.

\paragraph{Omitting all environmental components.}
Now suppose MoM fits only $\{K_g,I_N\}$.
Under $\operatorname{diag}(K_g)=\mathbf1$, this fitted set
satisfies the twin condition, so by Corollary~\ref{cor:omitted-diag} omission of $K_{nxe}=D^2$
contributes no bias to the additive genetic estimate.
However, the omission of the $G\times E$ kernel gives
\[
\boxed{
B_g^{(G)}
:=\E[\widehat\sigma_g^2\mid X,E]-\sigma_g^2
=\sigma_{gxe}^2
\frac{\langle\widetilde K_g,\widetilde K_{gxe}\rangle_F}
     {\|\widetilde K_g\|_F^2}.
}
\]
When both omitted variance components are positive and the
kernel inner product is nonzero, its sign is opposite to the
additive genetic bias in the $G+G\times E$ fit. Note that for this set of kernels and assumptions, $\hat\sigma^2_g$ coincides with the HE estimator.

\begin{proposition}[Vanishing environment-averaged bias]
\label{prop:gonly-exchangeable}
Under the assumptions above, suppose additionally that $E$
is conditionally exchangeable given $X$, with
$\mathbf1^\top E=0$ and $E^\top E=N$ almost surely.
For fixed $\sigma_{gxe}^2$,
\[
\boxed{
\E_E[B_g^{(G)}\mid X]
=-\frac{\sigma_{gxe}^2}{N-1}
\xrightarrow[N\to\infty]{}0.
}
\]
\end{proposition}

The proof is given in Supplementary Section \ref{sec:exchangeable-proof}.
Conditional exchangeability may be reasonable for
environments assigned independently of genotype, but can fail
when exposure varies with ancestry or relatedness, as may occur
in observational data. The conclusion concerns bias averaged
over environments, rather than every fixed realization.

\begin{figure}[t]
  \centering
  \includegraphics[width=\textwidth]{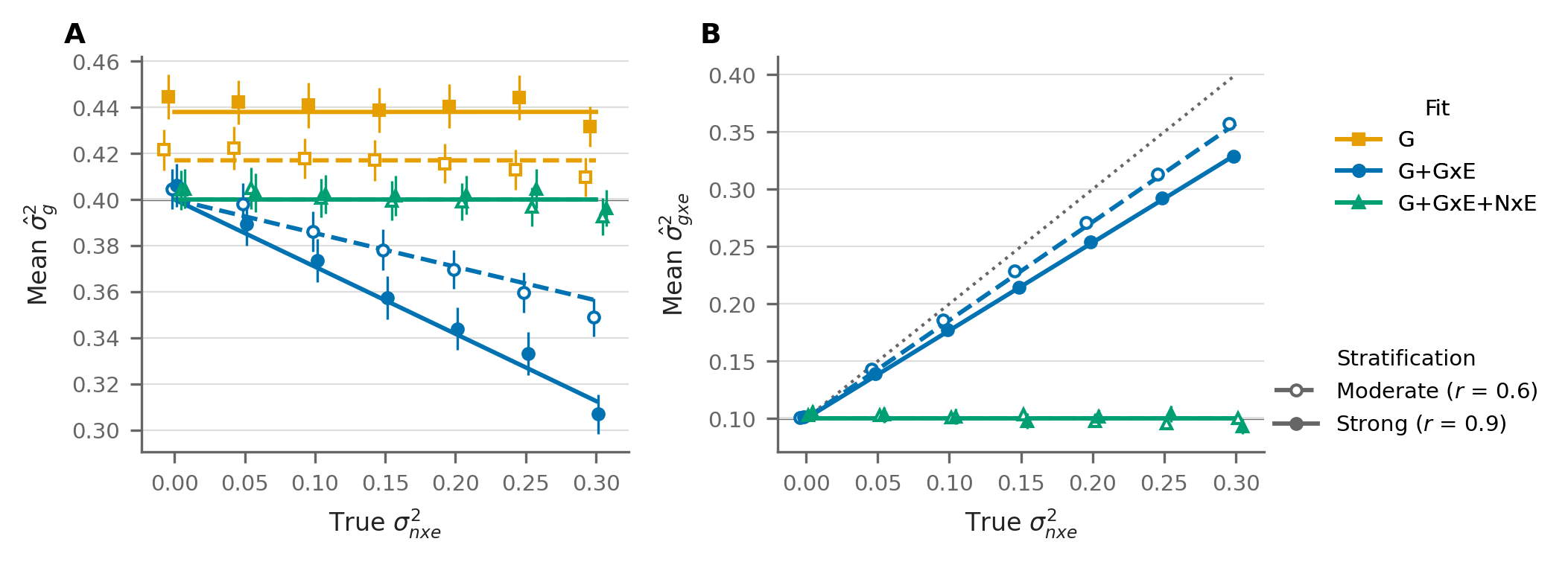}
  \caption{\textbf{Bias under one environmental variable.}
  Mean estimates of (A) $\sigma_g^2$ and (B) $\sigma_{gxe}^2$
  against the true $N\times E$ variance for three fitted models.
  All models include $I_N$. Grey horizontal lines mark the true
  values, $\sigma_g^2=0.4$ and $\sigma_{gxe}^2=0.1$;
  the dotted line in (B) marks
  $\sigma_{gxe}^2+\sigma_{nxe}^2$.
  Coloured lines show exact conditional expectations, and
  points show means over $2{,}000$ phenotype replicates with
  $95\%$ Monte Carlo confidence intervals.
  Dashed lines and open markers indicate moderate environmental
  stratification ($r=0.6$); solid lines and filled markers
  indicate strong stratification ($r=0.9$).
  The $G$-only model has no interaction estimate and is therefore
  shown only in (A).}
  \label{fig:exp1}
\end{figure}

\paragraph{Simulation results.}
Experiment~1 used $N=20{,}000$ simulated individuals from five demes with mild fine-scale structure and
$M=223{,}591$ LD-pruned SNPs. The environment was a noisy copy of the top PC of the GRM, with correlation $r=0.6$ or
$0.9$, so that it tracks ancestry as geographic exposures do. Phenotypes followed the $G+G\times E+N\times E$ model
with $\sigma^2_g=0.4$, $\sigma^2_{gxe}=0.1$ and $\sigma^2_{nxe}$ from $0$ to $0.3$ (detailed design in Supplementary
Section~\ref{sec:sim-design}).
Empirical means agree with the exact expectations in every setting (Figure~\ref{fig:exp1}). The full model recovers
both genetic components throughout. Omitting N$\times$E while retaining G$\times$E inflates
$\widehat\sigma^2_{gxe}$ linearly in $\sigma^2_{nxe}$. The predicted slope is $0.86$ at $r=0.6$ and $0.77$ at
$r=0.9$, below the bound of~$1$ from Proposition~\ref{prop:single-nxe-bias}. The same omission lowers
$\widehat\sigma^2_g$ (slopes $-0.15$ and $-0.29$). The $G$-only estimates do not change across the sweep, because
under the twin condition omitting N$\times$E does not affect the bias. They still exceed $\sigma^2_g$, because G$\times$E is
omitted. Stronger stratification increases both additive biases, but it decreases the interaction inflation. Closer
alignment of the environment with population structure therefore need not increase the bias of every component; the
effect depends on the joint geometry of the fitted and omitted kernels.

\paragraph{Extension to multiple environmental variables.}We extend the preceding model to $L$ standardized environmental
variables, allowing correlation between environments while
retaining independence between random-effect vectors.
Let $D_\ell=\operatorname{diag}(E_{:\ell})$,
$K_{gxe,\ell}=D_\ell K_gD_\ell$, and
$K_{nxe,\ell}=D_\ell^2$. Assume the full model
\[
y
=
X\beta
+
\sum_{l=1}^{L} (D_l X)\alpha_l
+
\sum_{l=1}^{L} D_l \delta_l
+
\varepsilon,
\]
with the true covariance
\[
\Sigma
=
\sigma_g^2K_g
+\sum_{\ell=1}^L
\sigma_{gxe,\ell}^2K_{gxe,\ell}
+\sum_{\ell=1}^L
\sigma_{nxe,\ell}^2D_\ell^2
+\sigma_e^2I_N.
\]
As before, assume $\operatorname{diag}(K_g)=\mathbf 1$.

Suppose we retain both the $G\times E$ and $N\times E$
components for environments in $\mathcal R$ and omit both
for environments in $\mathcal O=\{1,\ldots,L\}\setminus\mathcal R$,
while retaining $K_g$ and $I_N$.
The twin condition still holds, so omitted $N\times E$
components do not change or bias the fitted genetic components.

Let $\mathcal G=\{g\}\cup\{gxe,r:r\in\mathcal R\}$ index
the fitted genetic kernels, and let $T^{\mathrm{off}}$ be
their off-diagonal Gram matrix from
Theorem~\ref{thm:separation}.
Assuming the fitted kernels are linearly independent,
\begin{equation}
B_j
=
\sum_{\ell\in\mathcal O}\sigma_{gxe,\ell}^2
\left[(T^{\mathrm{off}})^{-1}
v^{\mathrm{off}}_\ell\right]_j,
\qquad j\in\mathcal G,
\label{eq:multiple-environment-bias}
\end{equation}
where $(v^{\mathrm{off}}_\ell)_j
=\langle\operatorname{off}(K_j),
\operatorname{off}(K_{gxe,\ell})\rangle_F$.
Thus, omitted interaction components contribute additively
to the bias, which distributes across the additive and
retained interaction components without a universal sign.
By Corollary~\ref{cor:h2-bias}, these signs and magnitudes also apply to
heritability pseudo-bias.

When all environmental components are omitted
($\mathcal R=\varnothing$), MoM fits only $\{K_g,I_N\}$,
and the additive genetic bias simplifies to
\[
B_g
=
\sum_{\ell=1}^L \sigma_{gxe,\ell}^2
\frac{\langle\widetilde K_g,\widetilde K_{gxe,\ell}\rangle_F}
     {\|\widetilde K_g\|_F^2},
\qquad
\widetilde K_{gxe,\ell}=K_{gxe,\ell}-I_N.
\]
These additive contributions may reinforce or cancel one another.
Omitted N$\times$E components contribute no bias,
since the fitted kernels satisfy the twin condition.

The environment-averaged vanishing bias result (Proposition \ref{prop:gonly-exchangeable}) applies separately to each omitted environmental variable.
For any subset $\mathcal A\subseteq\mathcal O$ whose environmental
vectors are individually exchangeable conditional on $X$ and the
retained environments, their combined contribution to the
environment-averaged additive bias is
$-\sum_{\ell\in\mathcal A}\sigma_{gxe,\ell}^2/(N-1)$,
while their contributions to the averaged biases of retained
interaction components are zero (Supplementary Section \ref{sec:multi-environment-average}).
Other omitted environmental variables may still contribute non-vanishing bias.

\subsection{Allowing for correlated random effects}
\label{sec:cross}
Most existing G$\times$E methods assume that the additive, interaction and noise effects are mutually independent. GxEMM~\cite{gxemm}, for example, describes, but does not fit or use in their analysis, correlations between the G$\times$E effects of different environments and between environment-specific noise terms; we allow all such correlations, including those of additive with interaction effects.
Let $D_0=I_N$. For SNP $j$ and individual $i$, let
$b_j=(\beta_j,\alpha_{1j},\ldots,\alpha_{Lj})^\top\sim(0,M^{-1}\Omega_G)$ and
$u_i=(\varepsilon_i,\delta_{1i},\ldots,\delta_{Li})^\top\sim(0,\Omega_E)$, all mutually independent, with
$\Omega_G,\Omega_E\succeq0$ indexed by $r,s=0,\ldots,L$. Then
\[
\Sigma=\sum_{r\le s}(\Omega_G)_{rs}K^G_{rs}+\sum_{r\le s}(\Omega_E)_{rs}K^E_{rs},
\qquad
K^G_{rs}=\begin{cases}D_rK_gD_r,&r=s,\\ D_rK_gD_s+D_sK_gD_r,&r<s,\end{cases}
\qquad
K^E_{rs}=\begin{cases}D_r^2,&r=s,\\ 2D_rD_s,&r<s.\end{cases}
\]
The diagonal entries of $\Omega_G$ and $\Omega_E$ are the variance components ($\sigma^2_g$, $\sigma^2_{gxe,\ell}$, $\sigma^2_e$, $\sigma^2_{nxe,\ell}$) and we call the off-diagonal entries
\emph{cross-covariances}: genetic, $\tau_{rs}=(\Omega_G)_{rs}$ (G to G$\times$E for $r=0$, and between the
G$\times$E effects of two environments for $r\ge1$), and residual, $\eta_{rs}=(\Omega_E)_{rs}$. Each
cross-covariance enters $\Sigma$ linearly through its own kernel. Most methods assume diagonal $\Omega_G$ and
$\Omega_E$ and therefore omit these kernels, and Theorem~\ref{thm:additive-bias} applies directly.

\paragraph{Residual cross-covariances.} Every $K^E_{rs}$ is diagonal. If the fitted set satisfies the twin
condition, Corollary~\ref{cor:omitted-diag} therefore implies that omitting any of the $\eta_{rs}$ leaves the
genetic estimates (here $\hat\sigma_g^2$, $\hat\sigma_{gxe,\ell}^2$, $\hat\tau_{rs}$) unchanged in every replicate, and by Corollary~\ref{cor:h2-bias} it leaves the heritability
estimates unchanged as well.
When $\operatorname{diag}(K_g)=\mathbf 1$, the twin condition requires only $I_N$ and $D_\ell^2$ for every fitted
$K_{gxe,\ell}$. The cross kernels $2D_rD_s$ are not needed.

\paragraph{Genetic cross-covariances.} The kernels $K^G_{rs}$ are not diagonal, so omitting them can bias the
estimates. Omitting a set $\mathcal C$ of genetic cross-covariances under the twin condition gives,
extending~\eqref{eq:multiple-environment-bias},
\[
B_{\mathcal G}=\sum_{(r,s)\in\mathcal C}\tau_{rs}\,(T^{\mathrm{off}})^{-1}v^{\mathrm{off}}_{rs},
\qquad
(v^{\mathrm{off}}_{rs})_a=\langle\operatorname{off}(G_a),\operatorname{off}(K^G_{rs})\rangle_F .
\]
These biases have no fixed sign. For example, in the $G$-only fit, an omitted G--G$\times$E covariance gives
$B_g=2\tau_{0\ell}\sum_{i\neq j}K_{g,ij}^2E_{j\ell}/\|\widetilde K_g\|_F^2$. This is the alignment of the
environment with each individual's total squared relatedness to the rest of the sample. For fixed $\tau_{0\ell}$, this bias vanishes on averaging
over environments.

\begin{figure}[t]
  \centering
  \includegraphics[width=\textwidth]{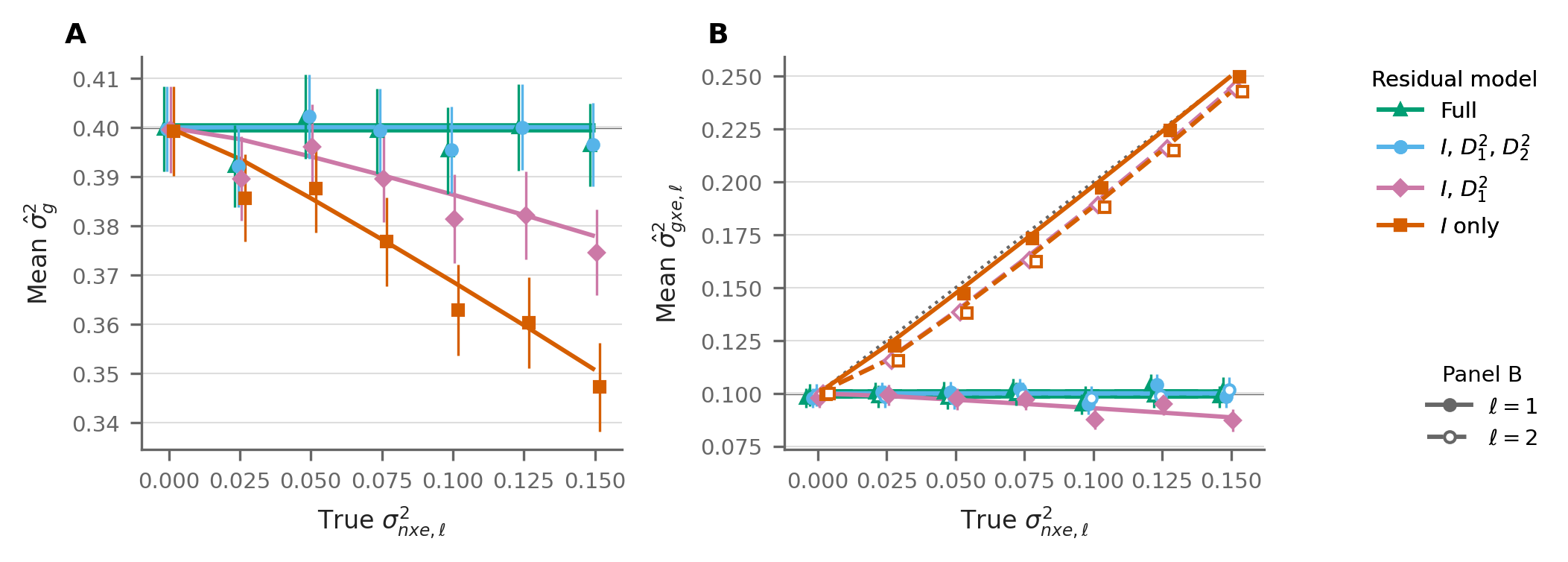}
    \caption{\textbf{Biased and unbiased estimates under two environmental variables with residual cross-covariances.}
  Mean estimates of (A) $\sigma^2_g$ and (B) $\sigma^2_{gxe,\ell}$ (solid/filled: $\ell = 1$; dashed/open:
  $\ell = 2$) against the true N$\times$E variance $\sigma^2_{nxe,1} = \sigma^2_{nxe,2}$, for four residual models.
  All fits contain $K_g$, $D_1K_gD_1$ and $D_2K_gD_2$. Full: $I$, $D_\ell^2$ and the residual cross-covariances
  $2D_\ell$, $2D_1D_2$ (correctly specified); $I, D_1^2, D_2^2$: cross-covariances omitted; $I, D_1^2$: in addition
  the N$\times$E of $E_2$ omitted; $I$ only: both N$\times$E terms omitted. Grey lines mark the true values
  ($\sigma^2_g = 0.4$, $\sigma^2_{gxe,\ell} = 0.1$); the dotted line in (B) is $\sigma^2_{gxe,1} + \sigma^2_{nxe}$.
  Lines: exact expected estimates; points: means over 2,000 replicates, drawn independently at each
  $\sigma^2_{nxe}$, with 95\% CIs.}
  \label{fig:exp2}
\end{figure}

\paragraph{Simulation results.} Experiment~2 used the genotypes of Experiment~1, two PC-aligned environments
($\operatorname{corr}(E_1,E_2)=0.25$), correlated residual terms and
$\sigma^2_{nxe,1}=\sigma^2_{nxe,2}=s\in[0,0.15]$ (Supplementary Section~\ref{sec:sim-design}). All fits include
the three genetic kernels and differ only in their residual kernels (Figure~\ref{fig:exp2}). The correctly specified fit and the fit omitting the residual cross-covariance kernels ($I,D_1^2,D_2^2$) both recover the genetic components.
In the second fit the twin condition holds only approximately, because $\operatorname{diag}(K_g)\approx\mathbf 1$,
and the remaining genetic bias is below $10^{-4}$. Removing $D_2^2$ breaks the twin of $D_2K_gD_2$. At $s=0.15$,
$\widehat\sigma^2_{gxe,2}$ then absorbs almost all of the omitted N$\times$E variance (bias $+0.144$), while
$\widehat\sigma^2_{gxe,1}$ is nearly unbiased ($-0.011$). The bias of $\widehat\sigma^2_g$ ($-0.022$) is about half
of its bias in the $I$-only fit ($-0.049$), where both interaction estimates are inflated ($+0.150$ and $+0.143$).
Each N$\times$E kernel thus protects the G$\times$E component of its own environment, and the biases of the omitted
terms add, as Theorem~\ref{thm:additive-bias} predicts. Empirical means agree with the exact expectations throughout.

\subsection{Genome partitioning and LD-dependent bias}
\label{sec:genomic-partition}

Variance-component models commonly partition SNPs by chromosome,
functional annotation, MAF or LD; multi-component methods such as
RHE-mc~\cite{rhemc} fit separate genetic kernels for these groups.
We study the bias caused by pooling groups with different per-SNP
effect variances into a single genome-wide component.

Partition the $M$ SNPs into disjoint sets $\mathcal S_1,\ldots,\mathcal S_R$
of sizes $M_r$, and let $X_r$ contain the corresponding genotype columns.
Assume each column is centered and has squared norm $N$. Define
\[
K_r=\frac{X_rX_r^\top}{M_r},
\qquad w_r=\frac{M_r}{M},
\qquad K_g=\sum_{r=1}^R w_rK_r,
\qquad
\Sigma=\sum_{r=1}^R\sigma_{g,r}^2K_r+\sigma_e^2I_N.
\]
Thus $\tr(K_r)=\tr(K_g)=N$, and the total genetic variance is
$\sigma_{g,\mathrm{tot}}^2=\sum_r\sigma_{g,r}^2$.
Suppose the unconstrained full-matrix MoM estimator fits only
$\{K_g,I_N\}$, with $K_g\ne I_N$.

\paragraph{Exact pooling bias.}
Writing $\widetilde K_r=K_r-I_N$ and
$\widetilde K_g=K_g-I_N$, eliminating the residual component
from the normal equations gives
\[
\E[\widehat\sigma_g^2\mid X]
=\sum_{r=1}^R\sigma_{g,r}^2
\frac{\langle\widetilde K_g,\widetilde K_r\rangle_F}
     {\|\widetilde K_g\|_F^2}.
\]
This expression has a direct LD interpretation. For SNP columns $x_j$,
define
\[
\ell_j=\sum_{k=1}^M\left(\frac{x_j^\top x_k}{N}\right)^2,
\qquad
\bar\ell_r=\frac{1}{M_r}\sum_{j\in\mathcal S_r}\ell_j,
\qquad
\bar\ell=\sum_{r=1}^R w_r\bar\ell_r.
\]
These are in-sample LD scores computed over all fitted SNPs,
including self-correlations and correlations between partitions.
Since $\tr(K_gK_r)=N^2\bar\ell_r/M$, the conditional bias is
\begin{equation}
\boxed{
B_g:=\E[\widehat\sigma_g^2\mid X]-\sigma_{g,\mathrm{tot}}^2
=
\frac{\sum_{r=1}^R\sigma_{g,r}^2(\bar\ell_r-\bar\ell)}
     {\bar\ell-M/N}.
}
\label{eq:genomic-pooling-bias}
\end{equation}
Proofs are provided in Supplementary
Section~\ref{app:genomic-partition}. The denominator is positive because
$\|\widetilde K_g\|_F^2=(N^2/M)(\bar\ell-M/N)>0$.
No unit-diagonal assumption is required.

Pooling therefore overestimates total genetic variance when the
genetic-variance-weighted mean LD score exceeds the SNP-weighted
mean, and underestimates it when the ordering is reversed.
The bias vanishes if all partitions have the same mean LD score,
or if $\sigma_{g,r}^2=w_r\sigma_{g,\mathrm{tot}}^2$ for every $r$,
corresponding to equal per-SNP effect variances.
By Corollary~\ref{cor:h2-bias}, these signs and magnitudes also apply to
heritability pseudo-bias.

\paragraph{Connection to literature.}
Min et al.~\cite{min2022} study LD-dependent bias under a
causal--noncausal SNP model with equal effect variances among
causal SNPs. Our analysis extends this setting to arbitrary
variance allocations across multiple genomic partitions and
provides an exact conditional bias formula for full-matrix MoM,
complementing their genotype-averaged approximation.
Their model is recovered as the two-partition special case
$B_g=\sigma_{g,\mathrm{tot}}^2
(\bar\ell_C-\bar\ell)/(\bar\ell-M/N)$,
where $\bar\ell_C$ is the mean LD score among causal SNPs.

\begin{figure}[t]
  \centering
  \includegraphics[width=\textwidth]{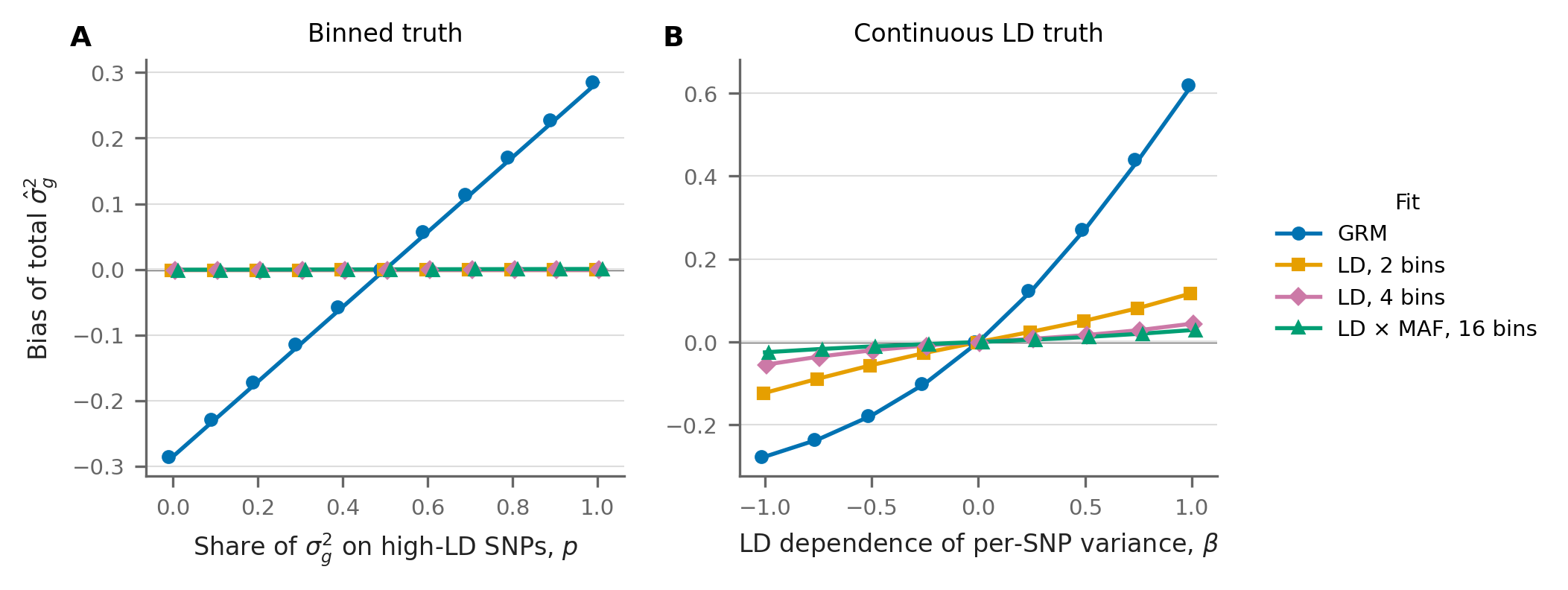}
  \caption{\textbf{Pooled MoM is biased when per-SNP heritability depends on LD; stratifying the GRM by LD
  removes the bias when the bins match the truth, and most of it otherwise.} Bias of the total genetic variance
  (true $\sigma^2_g = 0.4$) for a single GRM and for GRMs on 2 or 4 LD-score bins or 16 LD $\times$ MAF bins
  (GCTA-LDMS~\cite{yang2015}). Every SNP is causal. (A) Per-SNP variance is constant within each half of the SNPs (split at the
  median LD score), with a share $p$ of $\sigma^2_g$ on the high-LD half; the 2- and 4-bin fits contain this
  partition and are exactly unbiased, while the pooled bias is linear in $p$ (slope $0.57$). (B) Per-SNP variance
  $\propto \ell^\beta$, with $\ell$ the LD score; $\beta < 0$ puts more heritability per SNP on low-LD SNPs, and
  $\beta = 0$ is the uniform model, where every fit is unbiased. Lines: exact expected bias; points: means over
  2,000 replicates with 95\% CIs (mostly narrower than the markers).}
  \label{fig:exp3}
\end{figure}

\paragraph{Simulation results.}
Experiment~3 used $M=154{,}367$ SNPs on chromosomes 21 and 22, without LD pruning, with every SNP causal and
$\sigma^2_g=0.4$ (Supplementary Section~\ref{sec:sim-design}). When effect variances were constant within
each half of the SNPs split at the median LD score, pooled bias
varied linearly with the genetic variance allocated to the high-LD
half. Fits using two or four LD bins recovered the total genetic
variance because their partitions contained the true partition.
When per-SNP variances instead varied continuously as
$\ell_j^\beta$, LD and LD--MAF stratification reduced, but did not
generally eliminate, the bias. Means over $2{,}000$ phenotype
replicates agreed with the exact conditional predictions
(Figure~\ref{fig:exp3}).

\section{Summary and Future Work}

In this paper we presented a general formulation of the bias in MoM estimators of variance components and heritability in misspecified variance components models. The general formulation in Thm. \ref{thm:additive-bias} provides a framework for applying theoretical results like the diagonal separation of Thm. \ref{thm:separation}, and together they allow for drawing practical conclusions on estimation of gene-environment interaction effects (Sec. \ref{sec:single-environment}), genomic partitioning (Sec. \ref{sec:genomic-partition}) and more.
Our approach is also relevant to estimates from summary-statistics versions of MoM, such as LD-score regression and LDAK~\cite{speed2012}, which can be represented as MoM with the genetic inner product kernel modified by reweighting SNPs~\cite{zhou2017}. The validity of our specific results like the genome-partitioning bias of Sec. \ref{sec:genomic-partition} under this reweighting is a topic for future study.  \scomment{In these cases, it is simply a different set of moment equations. All MoM methods form moment equations of the $E(y^T A y)$ for some $A$. Currently, $A=K,I$ for additive. For summary stats, it is of the $A=W X^T$ but the principle is the same e.g. Zhou 2017 AoAS. Sumstats-based methods have further issues stemming from the mismatch in reference panels (use of external vs in-sample reference). These have been shown to introduce biases over and above the bias results here. Alternately, these results are a best-case scenario where in-sample LD is available.}

Several additional factors can affect the accuracy of MoM estimators, and we leave their detailed study for future work. 
When considering the accuracy of MoM estimators, estimation error or variance can play a role beyond the bias we are focused on. It is natural to argue that in big modern data problems with a small number of variance components, the variance of the estimators is expected to be small or even negligible, so bias may be the dominant factor. 
From a computational perspective, although MoM is more efficient than REML, it is often still impractical in Biobank-scale problems, and a common practical solution is to employ randomized MoM~\cite{rhe,rhemc}. This can introduce additional variance due to sampling and even additional bias when ratios like heritability are estimated.

\clearpage
\bibliographystyle{unsrt}
\bibliography{references}

@article{genie,
  author  = {Pazokitoroudi, Ali and Liu, Zhengtong and Dahl, Andrew and Zaitlen, Noah and Rosset, Saharon and Sankararaman, Sriram},
  title   = {A scalable and robust variance components method reveals insights into the architecture of gene-environment interactions underlying complex traits},
  journal = {The American Journal of Human Genetics},
  volume  = {111},
  number  = {7},
  pages   = {1462--1480},
  year    = {2024},
  doi     = {10.1016/j.ajhg.2024.05.015}
}

@article{gxemm,
  author  = {Dahl, Andy and Nguyen, Khiem and Cai, Na and Gandal, Michael J. and Flint, Jonathan and Zaitlen, Noah},
  title   = {A robust method uncovers significant context-specific heritability in diverse complex traits},
  journal = {The American Journal of Human Genetics},
  volume  = {106},
  number  = {1},
  pages   = {71--91},
  year    = {2020},
  doi     = {10.1016/j.ajhg.2019.11.015}
}

@article{gplemma,
  author  = {Kerin, Matthew and Marchini, Jonathan},
  title   = {A non-linear regression method for estimation of gene--environment heritability},
  journal = {Bioinformatics},
  volume  = {36},
  number  = {24},
  pages   = {5632--5639},
  year    = {2020},
  doi     = {10.1093/bioinformatics/btaa1079}
}

@article{stdpopsim,
  author  = {Adrion, Jeffrey R. and Cole, Christopher B. and Dukler, Noah and Galloway, Jared G. and Gladstein, Ariella L. and Gower, Graham and Kyriazis, Christopher C. and Ragsdale, Aaron P. and Tsambos, Georgia and Baumdicker, Franz and others},
  title   = {A community-maintained standard library of population genetic models},
  journal = {eLife},
  volume  = {9},
  pages   = {e54967},
  year    = {2020},
  doi     = {10.7554/eLife.54967}
}

@article{msprime,
  author  = {Baumdicker, Franz and Bisschop, Gertjan and Goldstein, Daniel and Gower, Graham and Ragsdale, Aaron P. and Tsambos, Georgia and Zhu, Sha and Eldon, Bjarki and Ellerman, E. Castedo and Galloway, Jared G. and others},
  title   = {Efficient ancestry and mutation simulation with msprime 1.0},
  journal = {Genetics},
  volume  = {220},
  number  = {3},
  pages   = {iyab229},
  year    = {2022},
  doi     = {10.1093/genetics/iyab229}
}

@article{leslie2015,
  author  = {Leslie, Stephen and Winney, Bruce and Hellenthal, Garrett and Davison, Dan and Boumertit, Abdelhamid and Day, Tammy and Hutnik, Katarzyna and Royrvik, Ellen C. and Cunliffe, Barry and Lawson, Daniel J. and Falush, Daniel and Freeman, Colin and Pirinen, Matti and Myers, Simon and Robinson, Mark and Donnelly, Peter and Bodmer, Walter},
  title   = {The fine-scale genetic structure of the {British} population},
  journal = {Nature},
  volume  = {519},
  number  = {7543},
  pages   = {309--314},
  year    = {2015},
  doi     = {10.1038/nature14230}
}

@article{haseman1972,
  author  = {Haseman, J. K. and Elston, R. C.},
  title   = {The investigation of linkage between a quantitative trait and a marker locus},
  journal = {Behavior Genetics},
  volume  = {2},
  number  = {1},
  pages   = {3--19},
  year    = {1972},
  doi     = {10.1007/BF01066731}
}

@article{ldsc,
  author  = {Bulik-Sullivan, Brendan K. and Loh, Po-Ru and Finucane, Hilary K. and Ripke, Stephan and Yang, Jian and {Schizophrenia Working Group of the Psychiatric Genomics Consortium} and Patterson, Nick and Daly, Mark J. and Price, Alkes L. and Neale, Benjamin M.},
  title   = {{LD} Score regression distinguishes confounding from polygenicity in genome-wide association studies},
  journal = {Nature Genetics},
  volume  = {47},
  number  = {3},
  pages   = {291--295},
  year    = {2015},
  doi     = {10.1038/ng.3211}
}

@article{tennessen2012,
  author  = {Tennessen, Jacob A. and Bigham, Abigail W. and O'Connor, Timothy D. and Fu, Wenqing and Kenny, Eimear E. and Gravel, Simon and McGee, Sean and Do, Ron and Liu, Xiaoming and Jun, Goo and others},
  title   = {Evolution and functional impact of rare coding variation from deep sequencing of human exomes},
  journal = {Science},
  volume  = {337},
  number  = {6090},
  pages   = {64--69},
  year    = {2012},
  doi     = {10.1126/science.1219240}
}

@article{zhou2017,
  author  = {Zhou, Xiang},
  title   = {A unified framework for variance component estimation with summary statistics in genome-wide association studies},
  journal = {The Annals of Applied Statistics},
  year    = {2017},
  volume  = {11},
  number  = {4},
  pages   = {2027--2051},
  doi     = {10.1214/17-AOAS1052}
}

@article{rhe,
  author  = {Wu, Yue and Sankararaman, Sriram},
  title   = {A scalable estimator of {SNP} heritability for biobank-scale data},
  journal = {Bioinformatics},
  year    = {2018},
  volume  = {34},
  number  = {13},
  pages   = {i187--i194},
  doi     = {10.1093/bioinformatics/bty253}
}

@article{rhemc,
  author  = {Pazokitoroudi, Ali and Wu, Yue and Burch, Kathryn S. and Hou, Kangcheng and Zhou, Aaron and Pasaniuc, Bogdan and Sankararaman, Sriram},
  title   = {Efficient variance components analysis across millions of genomes},
  journal = {Nature Communications},
  volume  = {11},
  pages   = {4020},
  year    = {2020},
  doi     = {10.1038/s41467-020-17576-9}
}

@article{hapmap2007,
  author  = {{International HapMap Consortium}},
  title   = {A second generation human haplotype map of over 3.1 million {SNPs}},
  journal = {Nature},
  volume  = {449},
  number  = {7164},
  pages   = {851--861},
  year    = {2007},
  doi     = {10.1038/nature06258}
}

@article{yang2015,
  author  = {Yang, Jian and Bakshi, Andrew and Zhu, Zhihong and Hemani, Gibran and Vinkhuyzen, Anna A. E. and Lee, Sang Hong and Robinson, Matthew R. and Perry, John R. B. and Nolte, Ilja M. and van Vliet-Ostaptchouk, Jana V. and others},
  title   = {Genetic variance estimation with imputed variants finds negligible missing heritability for human height and body mass index},
  journal = {Nature Genetics},
  volume  = {47},
  number  = {10},
  pages   = {1114--1120},
  year    = {2015},
  doi     = {10.1038/ng.3390}
}

@article{gcta2011,
  author  = {Yang, Jian and Lee, S. Hong and Goddard, Michael E. and Visscher, Peter M.},
  title   = {{GCTA}: a tool for genome-wide complex trait analysis},
  journal = {The American Journal of Human Genetics},
  volume  = {88},
  number  = {1},
  pages   = {76--82},
  year    = {2011},
  doi     = {10.1016/j.ajhg.2010.11.011}
}

@article{speed2012,
  author  = {Speed, Doug and Hemani, Gibran and Johnson, Michael R. and Balding, David J.},
  title   = {Improved heritability estimation from genome-wide {SNPs}},
  journal = {The American Journal of Human Genetics},
  volume  = {91},
  number  = {6},
  pages   = {1011--1021},
  year    = {2012},
  doi     = {10.1016/j.ajhg.2012.10.010}
}

@article{min2022,
  author  = {Min, Alan and Thompson, Elizabeth and Basu, Saonli},
  title   = {Comparing heritability estimators under alternative structures of linkage disequilibrium},
  journal = {G3: Genes, Genomes, Genetics},
  volume  = {12},
  number  = {8},
  pages   = {jkac134},
  year    = {2022},
  doi     = {10.1093/g3journal/jkac134}
}

@article{visscher2017,
  author  = {Visscher, Peter M. and Wray, Naomi R. and Zhang, Qian and Sklar, Pamela and McCarthy, Mark I. and Brown, Matthew A. and Yang, Jian},
  title   = {10 Years of {GWAS} Discovery: Biology, Function, and Translation},
  journal = {The American Journal of Human Genetics},
  volume  = {101},
  number  = {1},
  pages   = {5--22},
  year    = {2017},
  doi     = {10.1016/j.ajhg.2017.06.005}
}

@article{yang2010,
  author  = {Yang, Jian and Benyamin, Beben and McEvoy, Brian P. and Gordon, Scott and Henders, Anjali K. and Nyholt, Dale R. and Madden, Pamela A. and Heath, Andrew C. and Martin, Nicholas G. and Montgomery, Grant W. and Goddard, Michael E. and Visscher, Peter M.},
  title   = {Common {SNPs} explain a large proportion of the heritability for human height},
  journal = {Nature Genetics},
  volume  = {42},
  number  = {7},
  pages   = {565--569},
  year    = {2010},
  doi     = {10.1038/ng.608}
}

@article{finucane2015,
  author  = {Finucane, Hilary K. and Bulik-Sullivan, Brendan and Gusev, Alexander and Trynka, Gosia and Reshef, Yakir and Loh, Po-Ru and Anttila, Verneri and Xu, Han and Zang, Chongzhi and Farh, Kyle and others},
  title   = {Partitioning heritability by functional annotation using genome-wide association summary statistics},
  journal = {Nature Genetics},
  volume  = {47},
  number  = {11},
  pages   = {1228--1235},
  year    = {2015},
  doi     = {10.1038/ng.3404}
}

@article{robinson2017,
  author  = {Robinson, Matthew R. and English, Geoffrey and Moser, Gerhard and Lloyd-Jones, Luke R. and Triplett, Marcus A. and Zhu, Zhihong and Nolte, Ilja M. and van Vliet-Ostaptchouk, Jana V. and Snieder, Harold and {LifeLines Cohort Study} and others},
  title   = {Genotype--covariate interaction effects and the heritability of adult body mass index},
  journal = {Nature Genetics},
  volume  = {49},
  number  = {8},
  pages   = {1174--1181},
  year    = {2017},
  doi     = {10.1038/ng.3912}
}

@article{loh2015,
  author  = {Loh, Po-Ru and Bhatia, Gaurav and Gusev, Alexander and Finucane, Hilary K. and Bulik-Sullivan, Brendan K. and Pollack, Samuela J. and {Schizophrenia Working Group of the Psychiatric Genomics Consortium} and de Candia, Teresa R. and Lee, Sang Hong and Wray, Naomi R. and others},
  title   = {Contrasting genetic architectures of schizophrenia and other complex diseases using fast variance-components analysis},
  journal = {Nature Genetics},
  volume  = {47},
  number  = {12},
  pages   = {1385--1392},
  year    = {2015},
  doi     = {10.1038/ng.3431}
}

@article{speed2017,
  author  = {Speed, Doug and Cai, Na and {the UCLEB Consortium} and Johnson, Michael R. and Nejentsev, Sergey and Balding, David J.},
  title   = {Reevaluation of {SNP} heritability in complex human traits},
  journal = {Nature Genetics},
  volume  = {49},
  number  = {7},
  pages   = {986--992},
  year    = {2017},
  doi     = {10.1038/ng.3865}
}

@article{speed2019,
  author  = {Speed, Doug and Balding, David J.},
  title   = {{SumHer} better estimates the {SNP} heritability of complex traits from summary statistics},
  journal = {Nature Genetics},
  volume  = {51},
  number  = {2},
  pages   = {277--284},
  year    = {2019},
  doi     = {10.1038/s41588-018-0279-5}
}

@article{speed2020,
  author  = {Speed, Doug and Holmes, John and Balding, David J.},
  title   = {Evaluating and improving heritability models using summary statistics},
  journal = {Nature Genetics},
  volume  = {52},
  number  = {4},
  pages   = {458--462},
  year    = {2020},
  doi     = {10.1038/s41588-020-0600-y}
}

@article{gazal2017,
  author  = {Gazal, Steven and Finucane, Hilary K. and Furlotte, Nicholas A. and Loh, Po-Ru and Palamara, Pier Francesco and Liu, Xuanyao and Schoech, Armin and Bulik-Sullivan, Brendan and Neale, Benjamin M. and Gusev, Alexander and Price, Alkes L.},
  title   = {Linkage disequilibrium--dependent architecture of human complex traits shows action of negative selection},
  journal = {Nature Genetics},
  volume  = {49},
  number  = {10},
  pages   = {1421--1427},
  year    = {2017},
  doi     = {10.1038/ng.3954}
}

@article{evans2018,
  author  = {Evans, Luke M. and Tahmasbi, Rasool and Vrieze, Scott I. and Abecasis, Gon{\c{c}}alo R. and Das, Sayantan and Gazal, Steven and Bjelland, Douglas W. and de Candia, Teresa R. and {Haplotype Reference Consortium} and Goddard, Michael E. and others},
  title   = {Comparison of methods that use whole genome data to estimate the heritability and genetic architecture of complex traits},
  journal = {Nature Genetics},
  volume  = {50},
  number  = {5},
  pages   = {737--745},
  year    = {2018},
  doi     = {10.1038/s41588-018-0108-x}
}

@article{hou2019,
  author  = {Hou, Kangcheng and Burch, Kathryn S. and Majumdar, Arunabha and Shi, Huwenbo and Mancuso, Nicholas and Wu, Yue and Sankararaman, Sriram and Pasaniuc, Bogdan},
  title   = {Accurate estimation of {SNP}-heritability from biobank-scale data irrespective of genetic architecture},
  journal = {Nature Genetics},
  volume  = {51},
  number  = {8},
  pages   = {1244--1251},
  year    = {2019},
  doi     = {10.1038/s41588-019-0465-0}
}

@article{fastlmm,
  author  = {Lippert, Christoph and Listgarten, Jennifer and Liu, Ying and Kadie, Carl M. and Davidson, Robert I. and Heckerman, David},
  title   = {{FaST} linear mixed models for genome-wide association studies},
  journal = {Nature Methods},
  volume  = {8},
  number  = {10},
  pages   = {833--835},
  year    = {2011},
  doi     = {10.1038/nmeth.1681}
}

@article{gemma,
  author  = {Zhou, Xiang and Stephens, Matthew},
  title   = {Genome-wide efficient mixed-model analysis for association studies},
  journal = {Nature Genetics},
  volume  = {44},
  number  = {7},
  pages   = {821--824},
  year    = {2012},
  doi     = {10.1038/ng.2310}
}

@article{boltlmm,
  author  = {Loh, Po-Ru and Tucker, George and Bulik-Sullivan, Brendan K. and Vilhj{\'a}lmsson, Bjarni J. and Finucane, Hilary K. and Salem, Rany M. and Chasman, Daniel I. and Ridker, Paul M. and Neale, Benjamin M. and Berger, Bonnie and Patterson, Nick and Price, Alkes L.},
  title   = {Efficient {Bayesian} mixed-model analysis increases association power in large cohorts},
  journal = {Nature Genetics},
  volume  = {47},
  number  = {3},
  pages   = {284--290},
  year    = {2015},
  doi     = {10.1038/ng.3190}
}

\clearpage
\section*{Supplementary Material}

\setcounter{section}{0}
\renewcommand{\thesection}{S\arabic{section}}

\setcounter{equation}{0}
\renewcommand{\theequation}{S\arabic{equation}}

\setcounter{figure}{0}
\renewcommand{\thefigure}{S\arabic{figure}}

\setcounter{table}{0}
\renewcommand{\thetable}{S\arabic{table}}

\setcounter{proposition}{0}
\renewcommand{\theproposition}{S\arabic{proposition}}

Notation follows the main text. The true covariance is $\Sigma=\operatorname{Cov}(y\mid X,E)=\sum_{j\in J}\theta_jK_j$,
the fitted kernels are indexed by $\mathcal F\subseteq J$ and the omitted ones by $\mathcal O=J\setminus\mathcal F$,
$\widehat\theta_{\mathcal F}=T_{\mathcal F}^{-1}q_{\mathcal F}$ with $(T_{\mathcal F})_{rs}=\langle K_r,K_s\rangle_F=\tr(K_rK_s)$
and $(q_{\mathcal F})_r=y^\top K_ry$, and all expectations are conditional on $(X,E)$, so that every kernel and
$T_{\mathcal F}$ are fixed. We assume $\E[y\mid X,E]=0$ and that the fitted kernels are linearly independent.


\section{Unbiasedness under overspecification}
\label{sec:overspecification}

Here we allow the fitted index set to extend beyond $J$.
Let $\mathcal A$ index additional fixed symmetric kernels, with
$\mathcal A\cap J=\varnothing$, and fit
$\mathcal F=J\cup\mathcal A$.
Suppose $\E[y\mid X,E]=0$ and the expanded fitted kernels are
linearly independent. Then the exact, unconstrained MoM estimator
satisfies
\[
\E[\widehat\theta_j\mid X,E]=\theta_j
\quad (j\in J),
\qquad
\E[\widehat\theta_a\mid X,E]=0
\quad (a\in\mathcal A).
\]

\begin{proof}
Extend the true coefficient vector to $\mathcal F$ by setting
$\theta_a=0$ for every $a\in\mathcal A$. Then
\[
\Sigma=\sum_{s\in\mathcal F}\theta_sK_s.
\]
For each $r\in\mathcal F$,
\[
\E[y^\top K_r y\mid X,E]
=\tr(K_r\Sigma)
=\sum_{s\in\mathcal F}\theta_s\langle K_r,K_s\rangle_F.
\]
Collecting these identities gives
$\E[q_{\mathcal F}\mid X,E]=T_{\mathcal F}\theta_{\mathcal F}$.
Since $T_{\mathcal F}$ is fixed conditional on $(X,E)$ and
invertible by linear independence,
\[
\E[\widehat\theta_{\mathcal F}\mid X,E]
=T_{\mathcal F}^{-1}\E[q_{\mathcal F}\mid X,E]
=T_{\mathcal F}^{-1}T_{\mathcal F}\theta_{\mathcal F}
=\theta_{\mathcal F}.
\]
Thus the original coefficients remain unbiased, and each additional
coefficient has expectation zero.
\end{proof}

\section{Proof of Theorem~\ref{thm:separation} (diagonal--off-diagonal separation)}
\label{sec:separation-proof}

\begin{restated}{Theorem}{\ref{thm:separation}}
Suppose the fitted kernels $\mathcal F=\mathcal G\cup\mathcal H$ are linearly independent and satisfy the twin
condition, $\diagop(G_a)=\Phi\Lambda_{\cdot a}$ for every $G_a\in\mathcal G$. Then
$\widehat\theta_{\mathcal G}=(T^{\mathrm{off}})^{-1}q^{\mathrm{off}}$ and
$\widehat\theta_{\mathcal H}=(\Phi^\top\Phi)^{-1}\Phi^\top(y\circ y)-\Lambda\widehat\theta_{\mathcal G}$.
\end{restated}

\begin{proof}
The MoM estimator minimizes
\[
\Bigl\|yy^\top-\sum_{a\in\mathcal G}\theta_{\mathcal G,a}G_a-\sum_{h\in\mathcal H}\theta_{\mathcal H,h}H_h\Bigr\|_F^2
\]
over $(\theta_{\mathcal G},\theta_{\mathcal H})$. Every matrix splits as $A=\offop(A)+\diagop(A)$, and a diagonal matrix
is Frobenius-orthogonal to any matrix with zero diagonal, so $\|A\|_F^2=\|\offop(A)\|_F^2+\|\diagop(A)\|_F^2$. The
kernels in $\mathcal H$ are diagonal, so they contribute only to the diagonal part, and the objective decomposes as
\[
\underbrace{\Bigl\|\offop(yy^\top)-\sum_{a\in\mathcal G}\theta_{\mathcal G,a}\offop(G_a)\Bigr\|_F^2}_{\text{off-diagonal term}}
+\underbrace{\Bigl\|y\circ y-\sum_{a\in\mathcal G}\theta_{\mathcal G,a}\diagop(G_a)-\Phi\theta_{\mathcal H}\Bigr\|_2^2}_{\text{diagonal term}},
\]
where in the diagonal term we identify a diagonal matrix with the vector of its diagonal. By the twin condition,
$\sum_{a\in\mathcal G}\theta_{\mathcal G,a}\diagop(G_a)=\Phi\Lambda\theta_{\mathcal G}$, so the diagonal term equals
\[
\bigl\|y\circ y-\Phi(\Lambda\theta_{\mathcal G}+\theta_{\mathcal H})\bigr\|_2^2 .
\]
For fixed $\theta_{\mathcal G}$, this is a least-squares problem in $\theta_{\mathcal H}$ ($\Phi$ has full column rank
because the fitted kernels are linearly independent), minimized by
$\Lambda\theta_{\mathcal G}+\theta_{\mathcal H}=(\Phi^\top\Phi)^{-1}\Phi^\top(y\circ y)$, i.e.
\[
\theta_{\mathcal H}=(\Phi^\top\Phi)^{-1}\Phi^\top(y\circ y)-\Lambda\theta_{\mathcal G}.
\]
The minimized diagonal term is the squared norm of the residual of $y\circ y$ on $\Phi$, which does not depend on
$\theta_{\mathcal G}$. Hence $\widehat\theta_{\mathcal G}$ minimizes the off-diagonal term alone. Its normal equations are
\[
\sum_{b\in\mathcal G}\langle\offop(G_a),\offop(G_b)\rangle_F\,\theta_{\mathcal G,b}=\langle\offop(G_a),\offop(yy^\top)\rangle_F
=\sum_{n\neq m}(G_a)_{nm}y_ny_m,\qquad a\in\mathcal G,
\]
that is, $T^{\mathrm{off}}\widehat\theta_{\mathcal G}=q^{\mathrm{off}}$. $T^{\mathrm{off}}$ is invertible: if
$\sum_bc_b\offop(G_b)=0$, then $\sum_bc_bG_b=\Phi\Lambda c$ is a combination of the diagonal kernels, contradicting
linear independence of $\mathcal F$ unless $c=0$. Substituting $\widehat\theta_{\mathcal G}$ into the minimizer over
$\theta_{\mathcal H}$ gives the stated $\widehat\theta_{\mathcal H}$.
\end{proof}

\section{Proof of Corollary~\ref{cor:omitted-diag} (omitted diagonal kernels)}

\begin{restated}{Corollary}{\ref{cor:omitted-diag}}
Under the twin condition, any omitted component with a diagonal kernel induces zero bias in every fitted genetic
component.
\end{restated}

\begin{proof}
Let $\mathcal H'\supseteq\mathcal H$ contain only diagonal
kernels, and suppose both $\mathcal G\cup\mathcal H$ and
$\mathcal G\cup\mathcal H'$ are linearly independent.
Because the original fitted set satisfies the twin condition,
the diagonal of every $G_a\in\mathcal G$ lies in the span
of the kernels in $\mathcal H$, and hence also in the span
of those in $\mathcal H'$. Thus the expanded fitted set
satisfies the twin condition as well.

By Theorem~\ref{thm:separation}, the genetic estimates
in both fits equal
\[
\widehat\theta_{\mathcal G}
=(T^{\mathrm{off}})^{-1}q^{\mathrm{off}}.
\]
Neither $T^{\mathrm{off}}$ nor $q^{\mathrm{off}}$ depends
on the fitted diagonal kernels. Consequently,
\[
\widehat\theta_{\mathcal G}(\mathcal G\cup\mathcal H)
=
\widehat\theta_{\mathcal G}(\mathcal G\cup\mathcal H')
\]
for every phenotype vector $y$, regardless of its true
covariance model.

To establish the bias statement, let an omitted component
have diagonal kernel $K_j$. Its contribution to
$\E[q^{\mathrm{off}}\mid X,E]$ has entries
\[
\theta_j
\langle\offop(G_a),\offop(K_j)\rangle_F=0,
\qquad a\in\mathcal G.
\]
It therefore contributes zero to the bias of every fitted
genetic coefficient, even if other, nondiagonal components
are also omitted.

In particular, if all omitted kernels are diagonal, write
\[
\Sigma=\sum_{j\in\mathcal F}\theta_jK_j+\Delta,
\qquad \Delta \text{ diagonal}.
\]
Then
\[
\offop(\Sigma)
=\sum_{b\in\mathcal G}\theta_{\mathcal G,b}\offop(G_b).
\]
Using $\E[y\mid X,E]=0$, we obtain
\[
\begin{aligned}
\E[q^{\mathrm{off}}_a\mid X,E]
&=\sum_{n\ne m}(G_a)_{nm}\Sigma_{nm}\\
&=\langle\offop(G_a),\offop(\Sigma)\rangle_F\\
&=\sum_{b\in\mathcal G}
T^{\mathrm{off}}_{ab}\theta_{\mathcal G,b}.
\end{aligned}
\]
Hence
\[
\E[\widehat\theta_{\mathcal G}\mid X,E]
=(T^{\mathrm{off}})^{-1}
\E[q^{\mathrm{off}}\mid X,E]
=\theta_{\mathcal G}.
\]
Thus all fitted genetic coefficients are conditionally
unbiased when every omitted kernel is diagonal.
\end{proof}

\section{Proof of Lemma~\ref{lemma:trace} (trace conservation)}
\label{sec:trace-proof}
\begin{restated}{Lemma}{\ref{lemma:trace}}
Suppose $K_r=I_N$ for some $r\in\mathcal F$, $\Sigma=\sum_{j\in\mathcal F}\theta_jK_j+\Delta$ with
$\Delta=\sum_{j\in\mathcal O}\theta_jK_j$, and $B_j=\E[\widehat\theta_j\mid X,E]-\theta_j$. Then
$\sum_{j\in\mathcal F}\tr(K_j)B_j=\tr(\Delta)$, equivalently
$\sum_{j\in\mathcal F}\tr(K_j)\,\E[\widehat\theta_j\mid X,E]=\tr(\Sigma)$.
\end{restated}

The idea is that one of the normal equations of MoM, the one belonging to the residual kernel $I_N$, is a statement
about total variance: it forces the fitted model to reproduce the observed sum of squares $y^\top y$. Since this
equation is linear, it continues to hold in expectation, whatever the omitted covariance.

\begin{proof}
\emph{Step 1 (expected normal equations).} For any fixed symmetric $K$ and $\E[y]=0$,
$\E[y^\top Ky]=\E[\tr(Kyy^\top)]=\tr(K\Sigma)$. For a fitted kernel $K_r$, splitting $\Sigma$ into its fitted and
omitted parts gives
\[
\E[q_r]=\tr(K_r\Sigma)=\sum_{s\in\mathcal F}\tr(K_rK_s)\theta_s+\tr(K_r\Delta)
=(T_{\mathcal F}\theta_{\mathcal F})_r+v_r,
\qquad v_r:=\tr(K_r\Delta).
\]

\emph{Step 2 (the bias solves a linear system).} Since $T_{\mathcal F}$ is fixed and $\widehat\theta_{\mathcal F}$ is linear
in $q_{\mathcal F}$, $\E[\widehat\theta_{\mathcal F}]=T_{\mathcal F}^{-1}\E[q_{\mathcal F}]=\theta_{\mathcal F}+T_{\mathcal F}^{-1}v$.
Writing $B=(B_j)_{j\in\mathcal F}=\E[\widehat\theta_{\mathcal F}]-\theta_{\mathcal F}$, this says
\begin{equation}
\label{eq:TB}
T_{\mathcal F}B=v,\qquad\text{i.e.}\qquad\sum_{s\in\mathcal F}\tr(K_rK_s)B_s=\tr(K_r\Delta)\quad\text{for every }r\in\mathcal F.
\end{equation}
(This is Theorem~\ref{thm:additive-bias} multiplied through by $T_{\mathcal F}$.)

\emph{Step 3 (the row of $I_N$).} We do not need to invert $T_{\mathcal F}$; it suffices to read off one row of
\eqref{eq:TB}. Take $K_r=I_N$, which is fitted by assumption. Then $\tr(K_rK_s)=\tr(K_s)$ and
$\tr(K_r\Delta)=\tr(\Delta)$, so the $I_N$ row of \eqref{eq:TB} reads
\[
\sum_{s\in\mathcal F}\tr(K_s)B_s=\tr(\Delta),
\]
which is the first claim. Adding $\sum_{s\in\mathcal F}\tr(K_s)\theta_s=\tr\bigl(\sum_{s\in\mathcal F}\theta_sK_s\bigr)$
to both sides gives
\[
\sum_{s\in\mathcal F}\tr(K_s)\,\E[\widehat\theta_s]
=\tr\Bigl(\sum_{s\in\mathcal F}\theta_sK_s\Bigr)+\tr(\Delta)=\tr(\Sigma),
\]
which is the second.
\end{proof}

\paragraph{Replicate-wise version.} The same argument applied to the sample normal equations
$T_{\mathcal F}\widehat\theta_{\mathcal F}=q_{\mathcal F}$, whose $I_N$ row has right-hand side $q_{I_N}=y^\top I_Ny=y^\top y$,
gives, for every phenotype vector and without taking expectations,
\[
\sum_{s\in\mathcal F}\tr(K_s)\,\widehat\theta_s=y^\top y .
\]
Lemma~\ref{lemma:trace} is its expectation, since $\E[y^\top y]=\tr(\Sigma)$. The left side is the total variance
implied by the fitted model, summed over individuals; so whatever is omitted, the fitted components together always
account for exactly the observed variance, and misspecification only changes how it is divided among them.

\paragraph{Interpretation of $\tr(\Delta)$.} By the first identity, the omitted variance $\tr(\Delta)$ is
redistributed to the fitted components, component $j$ receiving $\tr(K_j)B_j$. If the omitted covariance has zero trace, the biases cancel
on the trace scale. For example,
$\tr(D_rK_g+K_gD_r)=2\sum_iE_{ir}=0$
when $\diagop(K_g)=\mathbf1$ and
$\mathbf1^\top E_{\cdot r}=0$, although this kernel
generally has a nonzero diagonal.

\section{Proof of Corollary~\ref{cor:h2-bias} (heritability pseudo-bias)}

\begin{restated}{Corollary}{\ref{cor:h2-bias}}
Suppose $K_r=I_N$ for some $r\in\mathcal F$, then $h^{2,*}_j-h^2_j=B_j\tr(K_j)/\tr(\Sigma)$ for every fitted genetic component $j$, with
$\operatorname{sgn}(h^{2,*}_j-h^2_j)=\operatorname{sgn}(B_j)$. Under the standard assumptions, the pseudo-bias equals $B_j$.
\end{restated}

\begin{proof}
By definition,
\[
h^2_j=\frac{\theta_j\tr(K_j)}{\tr(\Sigma)},
\qquad
h^{2,*}_j=\frac{(\theta_j+B_j)\tr(K_j)}{\sum_{r\in\mathcal F}(\theta_r+B_r)\tr(K_r)} .
\]
By Lemma~\ref{lemma:trace} the denominator of $h^{2,*}_j$ equals $\tr(\Sigma)$, so the two heritabilities share a
denominator and
\[
h^{2,*}_j-h^2_j=\frac{(\theta_j+B_j)\tr(K_j)-\theta_j\tr(K_j)}{\tr(\Sigma)}=\frac{\tr(K_j)}{\tr(\Sigma)}B_j .
\]
Since $\tr(K_j)>0$ for the genetic kernels ($K_g$ and $D_\ell K_gD_\ell$ are positive semidefinite and nonzero) and
$\tr(\Sigma)>0$, the sign of the pseudo-bias is that of $B_j$.

Under the standard assumptions, i.e.\ column-standardized genotypes with $\diagop(K_g)=\mathbf 1$, standardized
environmental variables ($E_{\cdot\ell}^\top E_{\cdot\ell}=N$), and unit average phenotypic variance
($\tr(\Sigma)=N$), we have
\[
\tr(K_g)=\sum_iK_{g,ii}=N,
\qquad
\tr(K_{gxe,\ell})=\sum_iK_{g,ii}E_{i\ell}^2=\sum_iE_{i\ell}^2=N,
\]
so $\tr(K_j)/\tr(\Sigma)=1$ and $h^{2,*}_j-h^2_j=B_j$ exactly.
\end{proof}

\section{Bias in $G\times E$ models with one environmental variable}
\label{sec:single-environment-proofs}

Throughout this section, $E$ is standardized:
$\mathbf1^\top E=0$ and $E^\top E=N$.
We assume $\operatorname{diag}(K_g)=\mathbf1$,
$\E[y\mid X,E]=0$, and linear independence of the fitted kernels.
The true covariance is
\[
\Sigma
=\sigma_g^2K_g+\sigma_{gxe}^2K_{gxe}
+\sigma_{nxe}^2D^2+\sigma_e^2I_N,
\qquad
D=\operatorname{diag}(E),\quad
K_{gxe}=DK_gD.
\]
Since both genetic kernels have trace $N$, their residuals
after Frobenius projection onto $I_N$ are
\[
\widetilde K_g=K_g-I_N,
\qquad
\widetilde K_{gxe}=K_{gxe}-I_N.
\]

\subsection{Omitting environment-specific noise}
\label{sec:omitted-nxe-proof}

\begin{proof}[Proof of Proposition~\ref{prop:single-nxe-bias}]
Suppose the fitted kernels are $\{K_g,K_{gxe},I_N\}$.
Eliminating the fitted $I_N$ coefficient from the normal
equations gives the following system for the genetic biases:
\[
\begin{pmatrix}
\|\widetilde K_g\|_F^2
&
\langle\widetilde K_g,\widetilde K_{gxe}\rangle_F\\
\langle\widetilde K_g,\widetilde K_{gxe}\rangle_F
&
\|\widetilde K_{gxe}\|_F^2
\end{pmatrix}
\begin{pmatrix}
B_g\\ B_{gxe}
\end{pmatrix}
=
\sigma_{nxe}^2
\begin{pmatrix}
\langle\widetilde K_g,D^2-I_N\rangle_F\\
\langle\widetilde K_{gxe},D^2-I_N\rangle_F
\end{pmatrix}.
\]
The unit-diagonal assumption implies
\[
\widetilde K_g=\operatorname{off}(K_g),
\qquad
\widetilde K_{gxe}
=\operatorname{off}(K_{gxe})+(D^2-I_N).
\]
By diagonal--off-diagonal orthogonality, the right-hand side is
\[
\sigma_{nxe}^2
\begin{pmatrix}
0\\ \|D^2-I_N\|_F^2
\end{pmatrix}.
\]
Linear independence of the fitted kernels implies that the
displayed Gram matrix is positive definite. Its first row gives
\[
B_g
=-\frac{\langle\widetilde K_g,\widetilde K_{gxe}\rangle_F}
        {\|\widetilde K_g\|_F^2}\,B_{gxe}.
\]
Substitution into the second row yields
\[
B_{gxe}
=\sigma_{nxe}^2
\frac{
\|\widetilde K_g\|_F^2\|D^2-I_N\|_F^2
}{
\|\widetilde K_g\|_F^2\|\widetilde K_{gxe}\|_F^2
-\langle\widetilde K_g,\widetilde K_{gxe}\rangle_F^2
}.
\]

To establish the upper bound, orthogonality and
Cauchy--Schwarz give
\[
\begin{aligned}
&\|\widetilde K_g\|_F^2\|\widetilde K_{gxe}\|_F^2
-\langle\widetilde K_g,\widetilde K_{gxe}\rangle_F^2\\
&\quad=
\|\widetilde K_g\|_F^2\|D^2-I_N\|_F^2
+\|\widetilde K_g\|_F^2
 \|\operatorname{off}(K_{gxe})\|_F^2
-\langle\widetilde K_g,\operatorname{off}(K_{gxe})\rangle_F^2\\
&\quad\geq
\|\widetilde K_g\|_F^2\|D^2-I_N\|_F^2.
\end{aligned}
\]
Consequently, if $\sigma_{nxe}^2>0$ and $D^2\neq I_N$,
then $0<B_{gxe}\leq\sigma_{nxe}^2$.
If either condition fails, the right-hand side of the bias
system is zero, so both genetic biases vanish.

Finally, since $(K_{gxe})_{ij}=K_{g,ij}E_iE_j$,
\[
\langle\widetilde K_g,\widetilde K_{gxe}\rangle_F
=\sum_{i\neq j}K_{g,ij}^2E_iE_j,
\]
which establishes the stated characterization of the
additive genetic bias.
\end{proof}

\subsection{The $G$-only fit}

\begin{proof}[Derivation of the $G$-only bias]
Suppose the fitted kernels are $\{K_g,I_N\}$.
After projecting out $I_N$, the genetic estimate is
\[
\widehat\sigma_g^2
=\frac{y^\top\widetilde K_g y}
       {\|\widetilde K_g\|_F^2}.
\]
Because $\widetilde K_g$ has zero diagonal,
\[
\langle\widetilde K_g,I_N\rangle_F
=\langle\widetilde K_g,D^2\rangle_F=0.
\]
Furthermore,
\[
\langle\widetilde K_g,K_g\rangle_F
=\|\widetilde K_g\|_F^2,
\qquad
\langle\widetilde K_g,K_{gxe}\rangle_F
=\langle\widetilde K_g,\widetilde K_{gxe}\rangle_F.
\]
Taking conditional expectations therefore gives
\[
\begin{aligned}
\E[\widehat\sigma_g^2\mid X,E]
&=\frac{\langle\widetilde K_g,\Sigma\rangle_F}
        {\|\widetilde K_g\|_F^2}\\
&=\sigma_g^2+
\sigma_{gxe}^2
\frac{\langle\widetilde K_g,\widetilde K_{gxe}\rangle_F}
     {\|\widetilde K_g\|_F^2}.
\end{aligned}
\]
Subtracting $\sigma_g^2$ proves the bias formula and shows
that the omitted $N\times E$ component contributes no bias.
\end{proof}

\subsection{Averaging over conditionally exchangeable environments}
\label{sec:exchangeable-proof}
\begin{proof}[Derivation of the environment-averaged bias]
Suppose the conditional distribution of $E$ given $X$ is
exchangeable, with $\mathbf1^\top E=0$ and $E^\top E=N$
almost surely. Exchangeability makes all conditional diagonal
second moments equal. Hence
\[
N\E[E_1^2\mid X]
=\E\left[\sum_{i=1}^NE_i^2\mid X\right]=N,
\]
so $\E[E_i^2\mid X]=1$ for every $i$.
Likewise, all conditional off-diagonal second moments are equal.
Using the centering condition,
\[
0
=\E\left[(\mathbf1^\top E)^2\mid X\right]
=N+N(N-1)\E[E_1E_2\mid X],
\]
and therefore
\[
\E[E_iE_j\mid X]=-\frac{1}{N-1},
\qquad i\neq j.
\]
Since $K_g$ is fixed given $X$,
\[
\begin{aligned}
\E_E[B_g^{(G)}\mid X]
&=\frac{\sigma_{gxe}^2}{\|\widetilde K_g\|_F^2}
  \sum_{i\neq j}K_{g,ij}^2\E[E_iE_j\mid X]\\
&=-\frac{\sigma_{gxe}^2}{N-1}
  \frac{\sum_{i\neq j}K_{g,ij}^2}
       {\|\widetilde K_g\|_F^2}\\
&=-\frac{\sigma_{gxe}^2}{N-1},
\end{aligned}
\]
where the last equality uses
$\widetilde K_g=\operatorname{off}(K_g)$.
For fixed $\sigma_{gxe}^2$, this quantity converges to zero
at rate $N^{-1}$.
\end{proof}
\section{Environment-averaged bias with multiple environmental variables}
\label{sec:multi-environment-average}

We use the multiple-environment setting of the main text:
$\operatorname{diag}(K_g)=\mathbf1$, and every environmental
vector is centered and standardized,
$\mathbf1^\top E_{\cdot\ell}=0$ and
$E_{\cdot\ell}^\top E_{\cdot\ell}=N$ almost surely.
The fit contains $K_g$, $I_N$, and both $K_{gxe,r}$ and $D_r^2$
for $r\in\mathcal R$; both components are omitted for
$\ell\in\mathcal O$.
Assume the fitted kernels are linearly independent and the
variance parameters are fixed.
Write $E_{\mathcal R}=(E_{\cdot r})_{r\in\mathcal R}$.
The fitted genetic kernels $G_a$ are indexed by
$\mathcal G=\{g\}\cup\{gxe,r:r\in\mathcal R\}$, and
$T^{\mathrm{off}}$ is their off-diagonal Gram matrix.

By \eqref{eq:multiple-environment-bias}, the conditional bias
decomposes as
\[
B_{\mathcal G}
=\sum_{\ell\in\mathcal O}b^{(\ell)},
\qquad
b^{(\ell)}
=\sigma^2_{gxe,\ell}(T^{\mathrm{off}})^{-1}
v^{\mathrm{off}}_\ell,
\qquad
(v^{\mathrm{off}}_\ell)_a
=\langle\offop(G_a),\offop(K_{gxe,\ell})\rangle_F.
\]
Omitted N$\times$E components contribute no genetic bias
because the fitted set satisfies the twin condition.

\begin{trivlist}
\item[\hskip\labelsep\bfseries Claim.]\itshape
Let $\mathcal A\subseteq\mathcal O$.
Suppose that, for each $\ell\in\mathcal A$, the conditional
distribution of $E_{\cdot\ell}$ given $(X,E_{\mathcal R})$
is exchangeable.
Then the combined bias contribution
$B_{\mathcal G}^{(\mathcal A)}
:=\sum_{\ell\in\mathcal A}b^{(\ell)}$ satisfies
\[
\E_E[B_g^{(\mathcal A)}\mid X,E_{\mathcal R}]
=-\frac{\sum_{\ell\in\mathcal A}\sigma^2_{gxe,\ell}}{N-1},
\qquad
\E_E[B_{gxe,r}^{(\mathcal A)}\mid X,E_{\mathcal R}]
=0
\quad(r\in\mathcal R).
\]
No independence between omitted environmental vectors is required.
\end{trivlist}

\begin{proof}
Fix $\ell\in\mathcal A$ and condition on $(X,E_{\mathcal R})$.
Exchangeability makes all diagonal second moments of
$E_{\cdot\ell}$ equal, and normalization gives
\[
\E[E_{i\ell}^2\mid X,E_{\mathcal R}]=1.
\]
Likewise, all off-diagonal second moments are equal.
Since $\mathbf1^\top E_{\cdot\ell}=0$ almost surely,
\[
0
=\E\left[
  \left(\sum_{i=1}^N E_{i\ell}\right)^2
  \,\middle|\,X,E_{\mathcal R}
\right]
=N+N(N-1)\E[E_{1\ell}E_{2\ell}\mid X,E_{\mathcal R}],
\]
so
\[
\E[E_{i\ell}E_{j\ell}\mid X,E_{\mathcal R}]
=-\frac{1}{N-1},
\qquad i\ne j.
\]
Because $(K_{gxe,\ell})_{ij}=K_{g,ij}E_{i\ell}E_{j\ell}$,
\[
\E[\offop(K_{gxe,\ell})\mid X,E_{\mathcal R}]
=-\frac{\offop(K_g)}{N-1}.
\]
Every fitted genetic kernel depends only on $X$ and
$E_{\mathcal R}$ and is therefore fixed under this conditioning.
Consequently,
\[
\begin{aligned}
\E[(v^{\mathrm{off}}_\ell)_a\mid X,E_{\mathcal R}]
&=\left\langle
  \offop(G_a),
  \E[\offop(K_{gxe,\ell})\mid X,E_{\mathcal R}]
  \right\rangle_F\\
&=-\frac{\langle\offop(G_a),\offop(K_g)\rangle_F}{N-1}\\
&=-\frac{T^{\mathrm{off}}_{a,g}}{N-1}.
\end{aligned}
\]
Let $e_g$ be the coordinate vector selecting the additive
component. In vector form,
\[
\E[v^{\mathrm{off}}_\ell\mid X,E_{\mathcal R}]
=-\frac{T^{\mathrm{off}}e_g}{N-1}.
\]
The twin condition and linear independence of the fitted
kernels imply that $T^{\mathrm{off}}$ is invertible.
Therefore,
\[
\begin{aligned}
\E_E[b^{(\ell)}\mid X,E_{\mathcal R}]
&=\sigma^2_{gxe,\ell}(T^{\mathrm{off}})^{-1}
  \E[v^{\mathrm{off}}_\ell\mid X,E_{\mathcal R}]\\
&=-\frac{\sigma^2_{gxe,\ell}}{N-1}e_g.
\end{aligned}
\]
Summing over $\ell\in\mathcal A$ proves the claim.
Only the conditional distribution of each individual
environment was used, so no independence assumption between
omitted environments is needed.
\end{proof}

If $\mathcal A=\mathcal O$, the claim describes the entire
environment-averaged genetic bias. Otherwise,
\[
\E_E[B_{\mathcal G}\mid X,E_{\mathcal R}]
=
-\frac{\sum_{\ell\in\mathcal A}\sigma^2_{gxe,\ell}}{N-1}e_g
+\sum_{\ell\in\mathcal O\setminus\mathcal A}
 \E_E[b^{(\ell)}\mid X,E_{\mathcal R}],
\]
and the remaining environments may contribute additional bias.
The contribution from $\mathcal A$ to the additive bias tends
to zero whenever
$\sum_{\ell\in\mathcal A}\sigma^2_{gxe,\ell}=o(N)$,
in particular when this total variance remains bounded.

When $\mathcal R=\varnothing$, conditioning on $X$ alone
suffices, and the result reduces to
Proposition~\ref{prop:gonly-exchangeable}, summed over
$\ell\in\mathcal A$. With retained environments, exchangeability
conditional on $X$ alone is generally insufficient.

\section{Simulation design}
\label{sec:sim-design}

\subsection{Genotypes}
\label{sec:genotypes}

\paragraph{Experiments 1 and 2.}
We simulated genotypes on the 22 autosomes with \texttt{stdpopsim} (v0.3.0)~\cite{stdpopsim} and \texttt{msprime}
(v1.4.2)~\cite{msprime}, under the European branch of the \texttt{OutOfAfrica\_2T12} demographic
model~\cite{tennessen2012} and the HapMapII (GRCh38) genetic maps~\cite{hapmap2007}. Fine-scale population structure
drives the biases we study, so we split the European population 200 generations ago into $K=5$ demes in a
stepping-stone chain, with migration rate $10^{-4}$ between adjacent demes. Each deme grew exponentially from
$N_e=8{,}000$ to $160{,}000$. The resulting pairwise $F_{ST}$ between demes, $\approx0.0017$, is in line with
estimates of regional structure in the UK~\cite{leslie2015}. We simulated $8{,}000$ individuals per deme and removed
SNPs with MAF $<1\%$. We then LD-pruned each chromosome greedily: a SNP was kept if its $r^2$ with each of the last
$300$ retained SNPs was below $0.1$. Finally, we thinned each chromosome evenly to a quota proportional to its
length, giving $M=223{,}591$ SNPs. We sampled $4{,}000$ individuals per deme, giving $N=20{,}000$.

\paragraph{Experiment 3.}
Because LD pruning reduces the variation in LD that Experiment 3 relies on, we simulated a second dataset for
chromosomes 21 and 22 under the same demographic model, genetic maps and deme structure, with $N=20{,}000$ individuals
($4{,}000$ per deme). We kept all biallelic SNPs with MAF $\ge1\%$ and performed no LD pruning, yielding $159{,}898$
SNPs. We then removed the $5{,}531$ SNPs in the chromosome 21 centromeric gap of the genetic map (10.65--12.97\,Mb,
where the map has zero recombination rate). With no recombination, the simulation produces a single 2.3\,Mb haplotype
block there, whereas SNPs in this region are largely absent from real genotype data. The final dataset has
$M=154{,}367$ SNPs. We computed the LD score of each SNP in-sample as the sum of its squared correlations with all
SNPs within 1\,cM (itself included), using the bias-corrected $\tilde r^2=r^2-(1-r^2)/(N-2)$~\cite{ldsc}. The median
LD score was 44.4 (5th--95th percentile 6.5--285.8).

In all experiments, genotypes were standardized in-sample to mean zero and unit variance, so that $\tr(K)=N$, and
held fixed across simulation replicates.

\subsection{Environments, phenotypes and fits}
\label{sec:phenotypes}

In every experiment we drew Gaussian phenotypes with $X$ and $E$ fixed, $2{,}000$ independent replicates per
setting, and fitted each model by exact MoM with no covariates. Expected estimates were computed with
Theorem~\ref{thm:additive-bias} from the realized kernels, without assuming a unit GRM diagonal
(Section~\ref{sec:kdiag}).

\paragraph{Experiment 1.} The environment was $E=\operatorname{std}(\lambda\,\mathrm{PC1}+z)$, where PC1 is the top
eigenvector of $K_g$, $z\sim N(0,I_N)$, and $\operatorname{std}$ centers the vector and scales it so that $E^\top E=N$.
PC1 is the leading axis of ancestry, so $E$ mimics an exposure that varies with ancestry, as geographic exposures do,
and is aligned with the off-diagonal structure of $K_g$ that drives the bias. We set $\lambda$ so that the realized
$\operatorname{corr}(E,\mathrm{PC1})$ was $r=0.6$ (moderate) or $0.9$ (strong stratification). Phenotypes followed the
$G+G\times E+N\times E$ model with $\sigma^2_g=0.4$, $\sigma^2_{gxe}=0.1$, $\sigma^2_{nxe}\in\{0,0.05,\ldots,0.3\}$ and
$\sigma^2_e=0.5-\sigma^2_{nxe}$, so that the
variance-component coefficients sum to one. We fitted the $G$, $G+G\times E$ and full models.

\paragraph{Experiment 2.} We kept the genotypes of Experiment~1 and built two environments in the same way: $E_1$
from PC1 and $E_2$ from $0.7\,\mathrm{PC1}+0.71\,\mathrm{PC2}$, each with correlation $0.6$ with its score, which
gives $\operatorname{corr}(E_1,E_2)=0.25$. We set $\Omega_G=\operatorname{diag}(0.4,0.1,0.1)$,
$\sigma^2_{nxe,1}=\sigma^2_{nxe,2}=s\in[0,0.15]$ and $\sigma^2_e=0.4-2s$. The residual correlations were $0.4$
between $\varepsilon$ and each $\delta_\ell$, and $0.5$ between $\delta_1$ and $\delta_2$. All fits contain $K_g$,
$D_1K_gD_1$ and $D_2K_gD_2$; the four residual models are listed in the caption of Figure~\ref{fig:exp2}.

\paragraph{Experiment 3.}
We set the total genetic variance to $\sigma_g^2=0.4$ and the residual variance to $\sigma_e^2=0.6$, so that the
total variance is $1$. Let $L_j$ denote the windowed, bias-corrected LD score defined in
Section~\ref{sec:genotypes}. These scores were used to construct the simulation architectures and fitted bins;
they differ from the all-SNP, uncorrected scores $\ell_j$ in the exact LD-score identities.

In the first design, SNPs were divided into low- and high-LD groups at the median of $L_j$, with group sizes
$M_L=77{,}184$ and $M_H=77{,}183$. For $p\in\{0,0.1,\ldots,1\}$, the per-SNP effect variances were
$(1-p)\sigma_g^2/M_L$ in the low-LD group and $p\sigma_g^2/M_H$ in the high-LD group. All SNPs had positive effect
variance for $0<p<1$; at either endpoint, one group had zero effect variance. This design is exactly the
partition model of the main text (Section~\ref{sec:genomic-partition}) with $R=2$: writing
$K_L=X_LX_L^\top/M_L$ and $K_H=X_HX_H^\top/M_H$, the true covariance is
$\Sigma=(1-p)\sigma_g^2K_L+p\sigma_g^2K_H+\sigma_e^2I_N$, so $\sigma^2_{g,L}=(1-p)\sigma_g^2$,
$\sigma^2_{g,H}=p\sigma_g^2$, and the pooled bias is given exactly by Eq.~\eqref{eq:genomic-pooling-bias}
(with the all-SNP scores $\ell_j$), linear in $p$. The second design is not of this form: the per-SNP variance
varies continuously, it is constant within no bin of any fit, and so every fitted partition is misspecified.
In the second design,
\[
\operatorname{Var}(\beta_j\mid X)
=\sigma_g^2\frac{L_j^\beta}{\sum_{k=1}^M L_k^\beta},
\qquad
\beta\in\{-1,-0.75,\ldots,1\}.
\]
Because the self-correlation contributes $1$ and the bias correction is small, every score was positive; the
smallest was $L_j=1.30$, so no score needed truncation before the powers were applied.

We fitted a single GRM, GRMs for two or four LD-score quantile bins, and GRMs for 16 LD--MAF
bins~\cite{yang2015}. Quantile bins had equal counts: the SNPs were ranked by $L_j$, ties were broken by genomic
position, and the SNP of rank $r\in\{0,\ldots,M-1\}$ was assigned to bin $\lfloor nr/M\rfloor$ of $n$, so bin sizes
differ by at most one. The LD--MAF fit first divided SNPs into four in-sample MAF groups, $[0.01,0.05)$,
$[0.05,0.1)$, $[0.1,0.2)$ and $[0.2,0.5]$ ($43{,}571$, $20{,}031$, $28{,}717$ and $62{,}048$ SNPs), and then into
LD-score quartiles defined within each MAF group, by the same rule. Each bin $b$ with $M_b$ SNPs had kernel
$K_b=X_bX_b^\top/M_b$, so that $\tr(K_b)=N$, and every fit also contained $I_N$. The total genetic variance of a
fit is the sum of its bin estimates. The two-bin fit used the same groups as the first simulation design, and the
four-bin partition, defined by the global quartiles of $L_j$, refined those groups. Thus both fits represented that
design's true covariance. The LD--MAF bins, defined within MAF groups, do not refine it. Exact expected estimates
were calculated from the realized covariance and fitted kernels as
\[
\E[\widehat\theta_{\mathcal F}\mid X]
=T_{\mathcal F}^{-1}
\bigl(\tr(K_r\Sigma)\bigr)_{r\in\mathcal F},
\]
rather than by substituting the windowed scores $L_j$ into the exact LD-score bias formula.

\subsection{Diagonal of the GRM}
\label{sec:kdiag}

\begin{figure}[h]
  \centering
  \includegraphics[width=0.5\textwidth]{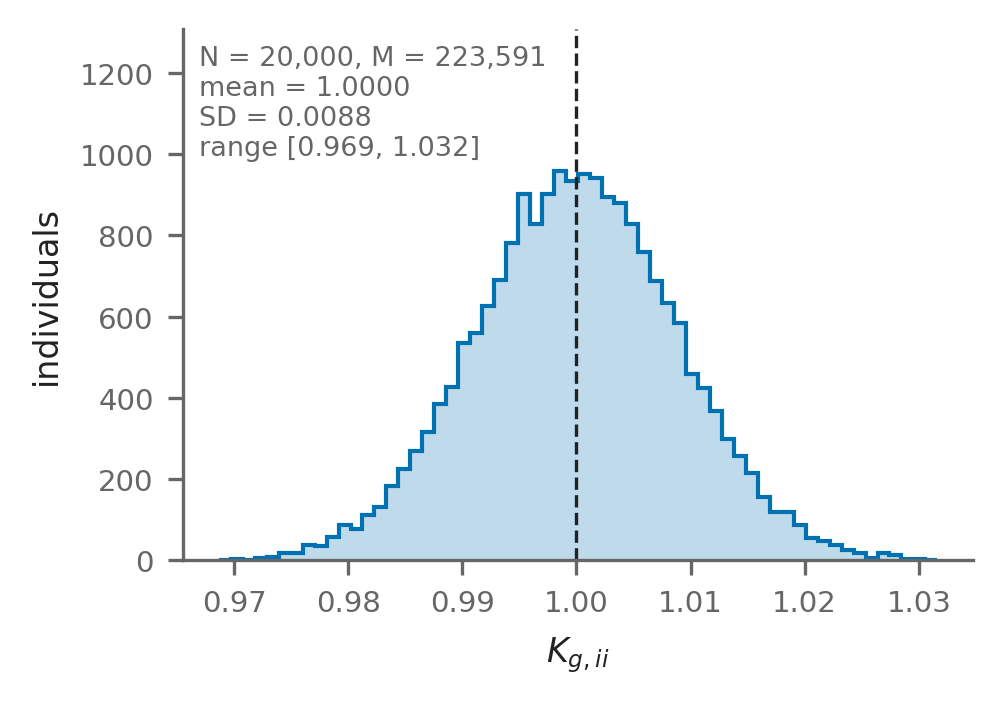}
  \caption{\textbf{The GRM diagonal is close to 1 under in-sample standardization.} Distribution of
  $K_{g,ii}=\frac1M\sum_{j=1}^Mx_{ij}^2$ over the $N=20{,}000$ individuals for the genotypes of Experiments 1 and 2
  (LD-pruned, 22 autosomes, $M=223{,}591$), with each SNP standardized in-sample to mean $0$ and variance $1$. The mean is
  exactly $1$ by construction ($\tr(K_g)=N$). The SD across individuals is $0.0088$, and every individual lies in
  $[0.969,1.032]$ ($98\%$ in $[0.979,1.021]$). There is no systematic difference between populations: the deme means
  range from $0.9989$ to $1.0005$, with within-deme SDs of $0.0087$--$0.0089$. The remaining deviation from $\mathbf 1$ has
  no detectable effect on our results.}
  \label{fig:supp-kdiag}
\end{figure}

The closed forms of the main text and the twin condition for $\{K_g,I_N\}$ assume $\diagop(K_g)=\mathbf 1$. With each SNP
standardized in-sample to mean $0$ and variance $1$, $K_{g,ii}=\frac1M\sum_{j=1}^Mx_{ij}^2$ has mean exactly $1$ over
individuals ($\tr(K_g)=N$), but the individual entries vary. Figure~\ref{fig:supp-kdiag} shows that on the genotypes of
Experiments 1 and 2 this variation is small. All theoretical expectations are nevertheless computed
from the realized kernels with Theorem~\ref{thm:additive-bias}, which does not require a unit diagonal.

\section{Proofs for genomic partitioning}
\label{app:genomic-partition}

Throughout this section, we condition on $X$ and assume
$\E[y\mid X]=0$.
Let $\mathcal S_1,\ldots,\mathcal S_R$ be a partition of the
$M$ SNPs into nonempty sets of sizes $M_r$.
Each genotype column $x_j$ is centered and satisfies
$x_j^\top x_j=N$.
Write
\[
K_r=\frac{X_rX_r^\top}{M_r},
\qquad
w_r=\frac{M_r}{M},
\qquad
K_g=\frac{XX^\top}{M}=\sum_{r=1}^R w_rK_r.
\]
The true conditional covariance and total genetic variance are
\[
\Sigma=\sum_{r=1}^R\sigma_{g,r}^2K_r+\sigma_e^2I_N,
\qquad
\sigma_{g,\mathrm{tot}}^2=\sum_{r=1}^R\sigma_{g,r}^2.
\]
Column normalization implies
\[
\tr(K_r)
=\frac{1}{M_r}\sum_{j\in\mathcal S_r}x_j^\top x_j=N,
\qquad
\tr(K_g)=N.
\]
Define $\widetilde K_r=K_r-I_N$ and
$\widetilde K_g=K_g-I_N$.
These matrices have trace zero, but need not have zero diagonal.

\subsection{Conditional expectation and bias of the pooled estimator}

\begin{proposition}
\label{prop:app-pooled-expectation}
Suppose $K_g\ne I_N$ and the fitted kernels are $\{K_g,I_N\}$.
Then
\[
\E[\widehat\sigma_g^2\mid X]
=\sum_{r=1}^R\sigma_{g,r}^2c_r,
\qquad
c_r=
\frac{\langle\widetilde K_g,\widetilde K_r\rangle_F}
     {\|\widetilde K_g\|_F^2}.
\]
Consequently,
\[
B_g:=\E[\widehat\sigma_g^2\mid X]
      -\sigma_{g,\mathrm{tot}}^2
=\sum_{r=1}^R\sigma_{g,r}^2(c_r-1).
\]
\end{proposition}

\begin{proof}
The normal equations for the unconstrained full-matrix MoM fit are
\[
\begin{pmatrix}
\tr(K_g^2)&N\\
N&N
\end{pmatrix}
\begin{pmatrix}
\widehat\sigma_g^2\\
\widehat\sigma_e^2
\end{pmatrix}
=
\begin{pmatrix}
y^\top K_gy\\
y^\top y
\end{pmatrix}.
\]
Subtracting the second equation from the first gives
\[
\bigl(\tr(K_g^2)-N\bigr)\widehat\sigma_g^2
=y^\top(K_g-I_N)y.
\]
Since $\tr(K_g)=N$,
\[
\tr(K_g^2)-N
=\tr\bigl((K_g-I_N)^2\bigr)
=\|\widetilde K_g\|_F^2>0.
\]
Thus the fitted kernels are linearly independent and
\[
\widehat\sigma_g^2
=\frac{y^\top\widetilde K_gy}
       {\|\widetilde K_g\|_F^2}.
\]
For any fixed symmetric matrix $A$, the conditional mean-zero
assumption gives
$\E[y^\top Ay\mid X]=\tr(A\Sigma)$.
Therefore,
\[
\E[\widehat\sigma_g^2\mid X]
=
\frac{
\sum_{r=1}^R\sigma_{g,r}^2\tr(\widetilde K_gK_r)
+\sigma_e^2\tr(\widetilde K_g)}
{\|\widetilde K_g\|_F^2}.
\]
The residual term vanishes because $\tr(\widetilde K_g)=0$.
Moreover,
\[
\tr(\widetilde K_gK_r)
=\tr\bigl(\widetilde K_g(K_r-I_N)\bigr)
=\langle\widetilde K_g,\widetilde K_r\rangle_F.
\]
This proves the expectation formula.
Subtracting $\sum_r\sigma_{g,r}^2$ proves the bias formula.
\end{proof}

\subsection{LD-score representation}

For the standardized genotype columns, define
\[
r_{jk}=\frac{x_j^\top x_k}{N},
\qquad
\ell_j=\sum_{k=1}^M r_{jk}^2,
\qquad
\bar\ell_r=\frac{1}{M_r}\sum_{j\in\mathcal S_r}\ell_j,
\qquad
\bar\ell=\frac{1}{M}\sum_{j=1}^M\ell_j
        =\sum_{r=1}^R w_r\bar\ell_r.
\]
The scores include all SNPs in the pooled kernel, including
self-correlations and correlations between different partitions.

\begin{proposition}
\label{prop:app-pooled-ld}
Under the assumptions of
Proposition~\ref{prop:app-pooled-expectation},
\[
c_r=\frac{\bar\ell_r-M/N}{\bar\ell-M/N},
\qquad
\bar\ell-M/N>0,
\]
and hence
\[
B_g
=\frac{\sum_{r=1}^R
       \sigma_{g,r}^2(\bar\ell_r-\bar\ell)}
       {\bar\ell-M/N}.
\]
\end{proposition}

\begin{proof}
Using $K_g=M^{-1}\sum_{k=1}^M x_kx_k^\top$ and
$K_r=M_r^{-1}\sum_{j\in\mathcal S_r}x_jx_j^\top$,
we obtain
\[
\begin{aligned}
\tr(K_gK_r)
&=\frac{1}{MM_r}
  \sum_{j\in\mathcal S_r}\sum_{k=1}^M
  \tr(x_kx_k^\top x_jx_j^\top)\\
&=\frac{1}{MM_r}
  \sum_{j\in\mathcal S_r}\sum_{k=1}^M
  (x_j^\top x_k)^2\\
&=\frac{N^2}{MM_r}
  \sum_{j\in\mathcal S_r}\ell_j
=\frac{N^2}{M}\bar\ell_r.
\end{aligned}
\]
Similarly,
\[
\tr(K_g^2)
=\frac{N^2}{M^2}\sum_{j=1}^M\ell_j
=\frac{N^2}{M}\bar\ell.
\]
Since both kernels have trace $N$,
\[
\begin{aligned}
\langle\widetilde K_g,\widetilde K_r\rangle_F
&=\tr(K_gK_r)-\tr(K_g)-\tr(K_r)+N\\
&=\frac{N^2}{M}\bar\ell_r-N,
\\
\|\widetilde K_g\|_F^2
&=\tr(K_g^2)-N
=\frac{N^2}{M}\left(\bar\ell-\frac{M}{N}\right).
\end{aligned}
\]
The latter expression is strictly positive because $K_g\ne I_N$.
Taking their ratio proves the expression for $c_r$.
Finally,
\[
c_r-1
=\frac{\bar\ell_r-\bar\ell}{\bar\ell-M/N},
\]
and substitution into
Proposition~\ref{prop:app-pooled-expectation}
proves the stated bias formula.
\end{proof}

\subsection{Bias direction and zero-bias cases}

\begin{corollary}
\label{cor:app-pooled-sign}
If $\sigma_{g,\mathrm{tot}}^2>0$, define
$p_r=\sigma_{g,r}^2/\sigma_{g,\mathrm{tot}}^2$.
Then
\[
\frac{B_g}{\sigma_{g,\mathrm{tot}}^2}
=
\frac{\sum_{r=1}^R p_r\bar\ell_r
      -\sum_{r=1}^R w_r\bar\ell_r}
     {\bar\ell-M/N}.
\]
Thus the sign of $B_g$ is the sign of the difference between
the genetic-variance-weighted and SNP-weighted mean LD scores.
In particular, $B_g=0$ if either
\[
\bar\ell_r=\bar\ell
\quad\text{for every }r,
\qquad\text{or}\qquad
\sigma_{g,r}^2=w_r\sigma_{g,\mathrm{tot}}^2
\quad\text{for every }r.
\]
These conditions are sufficient, but not necessary.
\end{corollary}

\begin{proof}
Divide the bias formula in
Proposition~\ref{prop:app-pooled-ld}
by $\sigma_{g,\mathrm{tot}}^2$ and use
$\sum_r p_r=1$ and $\bar\ell=\sum_r w_r\bar\ell_r$.
The denominator is strictly positive, proving the sign statement.

If all partition means equal $\bar\ell$, every term in the
bias numerator vanishes.
If instead $\sigma_{g,r}^2=w_r\sigma_{g,\mathrm{tot}}^2$,
the numerator equals
\[
\sigma_{g,\mathrm{tot}}^2
\left(\sum_r w_r\bar\ell_r-\bar\ell\sum_r w_r\right)=0.
\]
The latter condition corresponds to equal per-SNP effect
variances, since
$\sigma_{g,r}^2/M_r=\sigma_{g,\mathrm{tot}}^2/M$.
More generally, zero bias requires only
$\sum_r\sigma_{g,r}^2(\bar\ell_r-\bar\ell)=0$,
which can hold through cancellation without either sufficient
condition.
If $\sigma_{g,\mathrm{tot}}^2=0$, nonnegativity of the true
variance components implies $\sigma_{g,r}^2=0$ for every $r$,
and the bias is also zero.
\end{proof}

\subsection{The causal--noncausal special case}

Let $C$ contain $m$ causal SNPs and $F$ contain the remaining
$M-m$ SNPs, where $0<m<M$.
Assume equal per-SNP effect variances within $C$ and zero
effect variances within $F$, so that
\[
\Sigma=\sigma_{g,\mathrm{tot}}^2K_C+\sigma_e^2I_N,
\qquad
K_g=\frac{m}{M}K_C+\frac{M-m}{M}K_F.
\]
This is the two-partition case with variance allocation
$(\sigma_{g,\mathrm{tot}}^2,0)$.
Proposition~\ref{prop:app-pooled-ld} therefore gives
\[
B_g
=\sigma_{g,\mathrm{tot}}^2
\frac{\bar\ell_C-\bar\ell}{\bar\ell-M/N}.
\]
Furthermore,
\[
\bar\ell_C-\bar\ell
=\left(1-\frac{m}{M}\right)(\bar\ell_C-\bar\ell_F).
\]
For positive total genetic variance, the bias is therefore
positive, negative or zero according to whether
$\bar\ell_C$ is greater than, less than or equal to
$\bar\ell_F$.
If all SNPs are causal and have equal per-SNP effect variances,
the pooled covariance model is correctly specified and $B_g=0$.
Having all SNPs causal alone does not guarantee zero bias
when their effect variances differ.

\subsection{Heritability pseudo-bias}

Assume $\tr(\Sigma)>0$.
The true total genetic heritability is
\[
h_g^2=\frac{N\sigma_{g,\mathrm{tot}}^2}{\tr(\Sigma)}.
\]
The second normal equation in the pooled fit gives, for every $y$,
\[
N\bigl(\widehat\sigma_g^2+\widehat\sigma_e^2\bigr)=y^\top y.
\]
Taking conditional expectations yields
\[
N\left(
\E[\widehat\sigma_g^2\mid X]
+\E[\widehat\sigma_e^2\mid X]
\right)=\tr(\Sigma).
\]
Consequently, heritability evaluated at the expected fitted
coefficients satisfies
\[
h_g^{2,*}
=\frac{N\E[\widehat\sigma_g^2\mid X]}{\tr(\Sigma)},
\qquad
h_g^{2,*}-h_g^2
=\frac{N}{\tr(\Sigma)}B_g.
\]
Its sign therefore agrees with that of $B_g$, and it equals
$B_g$ when $\tr(\Sigma)=N$.
This establishes a pseudo-bias identity, not an identity for
$\E[\widehat h_g^2\mid X]-h_g^2$, because the plug-in
heritability has a random denominator.

\subsection{Fitting the true partition or a refinement}

\begin{proposition}
\label{prop:app-partition-refinement}
Suppose the fitted model includes $I_N$ and one normalized
genetic kernel for each set in the true partition or a
refinement of it. If the fitted kernels are linearly independent,
the sum of the fitted genetic variance estimates is conditionally
unbiased for $\sigma_{g,\mathrm{tot}}^2$.
\end{proposition}

\begin{proof}
Let $\mathcal A_1,\ldots,\mathcal A_Q$ be the fitted partition,
with $m_a=|\mathcal A_a|>0$ and
\[
H_a=\frac{X_{\mathcal A_a}X_{\mathcal A_a}^\top}{m_a}.
\]
Because this partition refines the true partition, each
$\mathcal A_a$ lies in a unique $\mathcal S_{r(a)}$.
For every true partition,
\[
K_r
=\sum_{a:r(a)=r}\frac{m_a}{M_r}H_a.
\]
Hence the true covariance can be written in the fitted model as
\[
\Sigma=\sum_{a=1}^Q\gamma_aH_a+\sigma_e^2I_N,
\qquad
\gamma_a=\frac{m_a}{M_{r(a)}}\sigma_{g,r(a)}^2.
\]
Let $T$ be the Frobenius Gram matrix of
$H_1,\ldots,H_Q,I_N$, let $q$ be the corresponding vector of
phenotypic quadratic forms, and write
$\vartheta=(\gamma_1,\ldots,\gamma_Q,\sigma_e^2)^\top$.
Then
\[
\E[q\mid X]=T\vartheta.
\]
Linear independence makes $T$ invertible, so
\[
\E[\widehat\vartheta\mid X]
=T^{-1}\E[q\mid X]=\vartheta.
\]
Finally,
\[
\begin{aligned}
\E\left[\sum_{a=1}^Q\widehat\gamma_a\mid X\right]
&=\sum_{a=1}^Q\gamma_a\\
&=\sum_{r=1}^R\sigma_{g,r}^2
  \sum_{a:r(a)=r}\frac{m_a}{M_r}\\
&=\sum_{r=1}^R\sigma_{g,r}^2
=\sigma_{g,\mathrm{tot}}^2.
\end{aligned}
\]
The true partition itself is the special case with no subdivision.
\end{proof}

This result applies to a refinement because the true covariance
remains in the fitted kernel span. An arbitrary LD-based partition
need not have this property when effect variances vary within its
bins, so stratification alone does not guarantee unbiasedness.
The proposition concerns the sum of variance-component estimates;
it also implies zero total heritability pseudo-bias, but does not
assert finite-sample unbiasedness of the plug-in heritability ratio.

\end{document}